\documentclass[amsmath,eqsecnum]{svjour2}
\usepackage{amsmath}
\usepackage{eufrak}
\usepackage[dvips]{graphicx}
\usepackage{bm}
\usepackage{color}

\newcommand{\rv}{{\vec r}}

\newcommand{\xv}{{\vec x}}

\newcommand{\Tr}{{\rm Tr}}

\newcommand{\nh}{{\hat{n}}}

\newcommand{\mm}[1]{{\bf #1}}

\def\mm#1{ {\mathbf #1}}
\begin{document}

\title{Topology and Geometry of Smectic Order on Compact Curved Substrates}
\author{Xiangjun Xing}
\institute{Department of Physics, Syracuse University, Syracuse, New York~13244, USA}
\email{xxing@physics.syr.edu}

\date{\today} 

\begin{abstract}
Smectic order on arbitrary curved substrate can be described by a differential form of rank one (1-form), whose geometric meaning is the differential of the local phase field of the density modulation.  The exterior derivative of 1-form is the local dislocation density.   Elastic deformations are described by superposition of exact differential forms.   We use the formalism of differential forms to systematically classify and characterize all low energy smectic states on torus as well as on sphere.   A two dimensional smectic order confined on either manifold exhibits many topologically distinct low energy states.  Different states are not accessible from each other by local fluctuations.   The total number of low energy states scales as the square root of the system area.  We also address the energetics of 2D smectic on a curved substrate and calculate the mean field phase diagram of smectic on a thin torus.  Finally, we discuss the motion of disclinations for spherical smectics as low energy excitations, and illustrate the interesting connection between spherical smectic and the theory of elliptic functions.   
 

\end{abstract}

\maketitle

\noindent

\section{Introduction}
\label{Sec:Intro}
The term smectic order~\cite{CMP:CL,LC:deGennes}, usually refers to one dimensional  modulation of certain physical quantity with well defined periodicity.  Being only partially ordered, smectic systems exhibit solid-like rigidity only in one dimension, together fluidity in the other dimensions.   As a consequence, their responses to external stress and fields are extremely anisotropic and exotic.   The physics of smectic order, including the anomalous elasticities resulting from thermal~\cite{GP1,GP2,smectic-KPZ-1,smectic-KPZ-2} and quenched fluctuations~\cite{RTaerogel1,RTaerogel2,RTaerogel6}, as well as nematic-Smectic transitions~\cite{PhysRevE.49.2964,PhysRevLett.32.292,PhysRevB.24.363}, have been subjected to intensive studies in the past few decades.   

Recent studies of crystalline order on sphere \cite{PhysRevLett.89.185502,PhysRevB.62.8738,A.R.Bausch03142003}, i.e. the generalized Thompson problem, have revealed rich and striking interplays between defects, topology, and geometry.  What is perhaps most surprising out of these studies is the fractionalization of $+1/6$ disclinations into grain boundaries of finite extension.  The appearance of these string-like objects is purely due to the interaction between elastic strain and Gaussian curvature, and has no analogue in flat space.   More generally, the connection between defects, geometry and orders has been one of the man theme of soft condensed matter physics in the past two decades~\cite{CMP:Nelson}.

%

The problem of smectic order on curved surfaces have attracted substantial attention from theorists and experimentalists recently.   For example, Santangelo et. al.~\cite{santangelo:017801} have analyzed the energetics of smectic phase formed by block copolymer films coated on a curved substrate.   Coating of colloidal particles with diblock copolymer films may also provide alternative mechanism for self-assembling, which can be further exploited for designs of novel artificial atoms~\cite{Nelson-tetravalent}.   Kleman and Blanc~\cite{Kleman-smectic-confinement-01} studied a colloidal particle immersed in a smectic liquid crystal.  If the boundary conditions are such that the nematic order must be tangent to the boundary, the smectic layers intersect the surface vertically.   The loci of intersection then naturally define a stack of equal-distance lines, i.e. two dimensional smectic order, on the boundary surface.  Smectic orders are also believed to arise when a flexible charged polymer is adsorbed onto an oppositely charged curved surface, such as that of a colloidal particle~
\cite{RevModPhys.74.329,Park-overcharging-EPL-99,Mateescu-overcharging-EPL-99,NetzR.R._ma990263h,NetzR.R._ma990264+}.  As a consequence of the electrostatic attraction from the substrate, as well as the repulsion between different chain segments, the polymer may form equal-distanced layer pattern on the surface.   More recently, polymer vesicles with in-plane smectic order has been synthesized~\cite{Li-Smectic-Vesicle}, which provides an experimental realization of smectic order on two-dimensional flexible membranes.   Last but not least, motivated by structures formations in biological systems, stripe patterns on sphere were simulated via reaction-diffusion systems~\cite{PhysRevE.60.4588,PhysRevE.67.036206}.   

In the two dimensional flat space, dislocations of smectic orders, cost only finite energy, and hence always destroy the long range smectic order at any finite temperature.  Nevertheless, our primary interest in this work is smectic orders on compact, i.e. closed,  substrates that are of finite size and have no boundary, such as torus and sphere.  Typical examples include liquid film with in-plane modulation on the surface of a colloidal particle at the micron scale, and a polymer vesicle with in-plane stripe pattern.  We are therefore free of taking any thermodynamic limit.  In sufficiently low temperature, then, proliferation of dislocations is strongly suppressed, and the relevant states are the low energy states with {\em minimal number of defects}.  A question of primary importance is the existence of defects-free smectic states on a given substrate topology. If the answer is negative, what do minimally defected state look like?  More generally, we are interested in classifying all low energy, minimally-defected states on topological basis, and finding their connections with the substrate topology.  The answer to the above questions are of topological significance, and hence apply to smectic order both on rigid curved substrates and on flexible membranes.  Simultaneously, we are also interested in the geometric as well as energetic aspects of low energy smectic states on {\em rigid}, curved substrates \footnote{We
note that some interesting results on the interplay between smectic order and extrinsic as well as intrinsic curvature have been explored by Santangelo {\it et al} recently~\cite{santangelo:017801}.   }.   The energetics of smectic order on flexible membranes is more complicated and will be addressed in a separate publication \cite{polymer-vesicle-unpublished}.  In the exploration of these topological and geometric questions, we shall find the modern formalism of differential forms especially helpful.

A few generic results about smectic order on compact substrates emerge from our study.  Firstly, smectic orders both on sphere and torus exhibit many topologically distinct and minimally defected low energy states.  These states can be classified by two integer topological charges, which are related to the substrate topology and to the disclination pattern.   They are approximately degenerate to each other, with the energy differences scaling sublinearly with the system size.   The total number of these low energy states scales as the square root of the system size.   For a smectic on sphere, twisting of closely bound disclination pairs constitutes an interesting mode of low energy excitation that allows us to prove the topological equivalence of many seemingly very different low energy states.  For smectic on a torus, we study the energetic couplings between smectic order and extrinsic curvature in more detail.   We calculate the mean field phase diagram and find a continuous chiral-achiral transition that is driven by the competition between elastic strain energy and the energetic coupling between director field and the extrinsic curvature.   Some parts of our results were already published elsewhere \cite{Xing-curved-smectic-short}.

There are two complementary descriptions of smectic ordering.   In the weakly ordered regime, a smectic order can be described as one dimensional modulation of the relevant physical quantity with preferred wave length $d_0$:  
\begin{eqnarray}
\rho(x) = \rho_0 + \rho_1 \, \cos \Theta(x),
\label{density-0}
\end{eqnarray}
where $\Theta(x)$ is the phase field of the density modulation.  In the flat space and using Cartesian coordinate system, the ground state is characterized by a phase field 
\begin{eqnarray}
\Theta(x) = q_0 \,\nh \cdot \xv, 
\label{Theta-ground}
\end{eqnarray}
with $q_0 = 2 \pi/d_0$ the wave vector, and the unit vector $\nh$ the direction of modulation.  In the ground state Eq.~(\ref{Theta-ground}) in the flat space,  $\Theta(x)$ is a globally well-defined field.  This is no longer true in the presence of dislocations, or when the system is confined in a closed surface.   Moreover, the density profile Eq.~(\ref{density-0}) is invariant under the transformation $\Theta(x) \rightarrow- \Theta(x)$, i.e.  the unit vector $\nh$ is defined only up to a global sign.  This fact is intimately related to the $\pi$-rotation symmetry of the underlying nematic order, and leads to the existence of $\pm 1/2$ disclinations.


In the strongly ordered regime, a purely geometric description is more economical, where the smectic order is characterized by a stack of mathematical, i.e. infinitely thin, surfaces, which correspond to the ridges of the density profile Eq.~(\ref{density-0}).  In a $d$ dimensional space, each surface, called a smectic layer, is a $d-1$ dimensional submanifold.  In the flat space and in the absence of disloation, it is possible to assign a distinct phase $2k\pi$ to each layer, where $k$ is an integer.   For example, in the ground state Eq.~(\ref{Theta-ground}), each layer is a plane given by $q_0 \nh \cdot \vec{x} = 2 k \pi$; the separation between neighboring layers is uniformly $d_0 = 2 \pi/q_0$.   Elastic deformations can be described by continuous, single valued phonon fields, defined relative to the ground state.  Dislocations are traditionally described, again relative to ground state by singular, multiple-valued displacement fields that are properly quantized~\cite{CMP:CL}.  Equivalently, the phase field $\Theta(x)$ in Eq.~(\ref{density-0}) becomes multiple-valued in the presence of dislocations.



\begin{figure}
\begin{center}
\includegraphics[width=4.5cm]{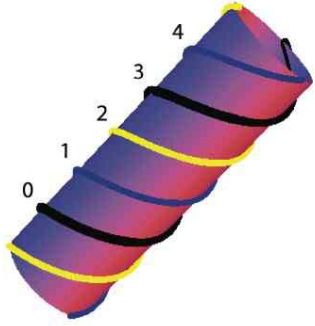}
\caption{(Color online) We label different layers by integer $n = 0, 1, 2, \ldots$ on a cylinder.  As the thick layer $0$ (black) wraps around the cylinder once and comes back, however it becomes the layer $3$.  This integer $3$ behaves as a topological charge and classify all low energy smectic states on a cylinder with no local defects.  This global topological charge is stable against all local fluctuations. }
\label{close-loop-cylinder}
\end{center}
\end{figure}

On a compact substrate, the phase field $\Theta(x)$ may be multiple-valued even in the absence of defects.  Careful analysis of this strange possibility leads to characterization of all low energy smectic states with minimal number of defects on closed surfaces.   As a simple example, let us consider a defects-free smectic state on a cylinder~\footnote{A cylinder is not a compact manifold such as torus or sphere.  It however serves as the simplest example that exhibit many topologically distinct low energy smectic states. }.  As shown in Fig.~\ref{close-loop-cylinder}, we locally identify a layer and label it with integer $0$.  We shall also chose a positive direction and label all neighboring layers successively as $-1,-2,\ldots$ and $1,2,\ldots$.  As shown in Fig.~\ref{close-loop-cylinder}, as the 0-th layer wraps around the cylinder once along the right hand direction, it comes back as the $3$-rd layer.   The phase field $\Theta$ is therefore well defined only up to $6 \pi$.   In a more general case, clearly, the phase uncertainty is $2 k \pi$, where $k$ is an arbitrary integer.   The phase field $\Theta(x)$ is single-valued oly if $k = 0$.  It is easy to see that the integer $k$ is also the minimal number of smectic layers that one has to cut in order to cut the cylinder into two halves in the direction perpendicular to the axis.  

\begin{figure}
\begin{center}
\includegraphics[width=5.5cm]{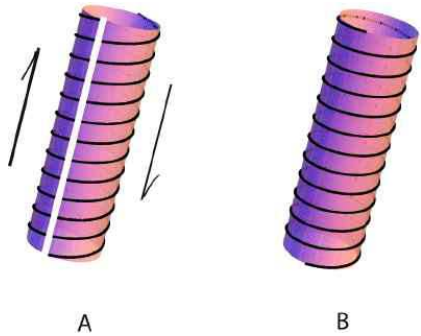}
\caption{How to change the global dislocation charge $k$: A) Starting from a state with $k=2$, cut the whole cylinder open along the cylinder axis. B) Shear the cylinder and reconnect all broken layers in a different way.  The final state is $k=1$.  Equivalently one can also generate a pair of dislocations (layer terminations), by cutting one layer,  and let them propagate in opposite direction through the whole cylinder.  Either way results in a global rearrangement of the smectic layer pattern. }
\label{cylinder-cut}
\end{center}
\vspace{-5mm}
\end{figure}

Note that two states with different $k$'s are not continuously accessible from each other.  
For such a purpose, a global and discontinuous rearrangement is needed.   For example, one may cut all layers along the direction of cylinder axis (Fig.~\ref{cylinder-cut}A) and reconnect them in a different way (Fig.~\ref{cylinder-cut}B), or, equivalently, one may generate a pair of dislocations (by cutting one layer) and move them along opposite direction out of the system.   In each process, one has to cut infinite number of smectic layers.   Therefore on a cylinder there exist {\em an infinite number of defects-free smectic states that are topologically distinct from each other}, each characterized by an integer $k$.   

As will be discussed in detail in Sec.~\ref{Sec:energetics}, the total free energy of smectic order on curved substrate consists of the energy cost for smectic layer compression/dilation and the layer bending energy cost.   It is clear that for all states the layer spacing can be continuously tuned to be $d_0$, the energetically preferred value.  Therefore the layer compression/dilation energy cost vanishes for every $k$.  The bending energy cost, on the other hand, is different for each $k$, and generically depends on several elastic moduli.  Nevertheless, its order of magnitude can be estimated as 
$f_{\rm bending} =  F_{\rm bending} /{A} \sim K /{R^2}, $
where $L$ and $R$ are the height and radius of the cylinder respectively.   The free energy difference per unit area of all states with different integer $k$ therefore scales inversely with $R^2$:  {\em all these topologically distinct smectic states on a cylinder become approximately degenerate as the system size becomes large comparing to the layer spacing. }

An integer that classifies degenerate but topologically distinct ground states is usually called a {\em topological quantum number}~\cite{Touless-TQN,XGWen}.  We shall see that the integer $k$ in the present case also shares essential similarities with the Burgers' number of a dislocation, but originates from the global (i.e. topological) structure of the underlying substrate.  It therefore may also be called a {\em global dislocation charge}.   As we shall show in great detail in this work, the existence of many topological distinct low energy states is a generic property of smectic on compact substrates.   In contrast, all defects free smectic states in flat space are topologically identical to each other.     

It is important to note that crystalline orders on cylinder has been intensively studied in the context of carbon nanotube~\cite{PhysRevLett.68.1579,saito:2204}.  It was shown that a cylindrical carbon nanotube is characterized by {\em two} integer charges, since a two dimensional solid has two independent density modulations.   The intuitive, geometric approach~\cite{PhysRevLett.68.1579,saito:2204} that leads to this result is to cut and roll a sheet of graphite into a tube with no defects.  This method provides a general and intuitive way of constructing closed-but-not-exact differential form on curved surface.  Essentially the same method is also used in the construction of nontrivial fiber bundle structures~\cite{book:Frankel,book:Nakahara}.


The remaining of this paper is organized as follows.  In Sec.~\ref{Sec:Forms} we introduce differential forms as a natural description of one dimensional translational order on curved surface.  We also discuss cohomology theory and apply it to smectic order on a torus.   In Sec.~\ref{Sec:heuristic} we give a more in-depth discussion of various topological aspects for smectic systems on curved substrates, as well as explore the relation between elastic deformations and exact differential forms.  In Sec.~\ref{Sec:energetics}, we discuss the energetic aspect of smectic orders in curved space, and calculate the ground state phase diagram of defects-free smectic order on a thin torus.   In Sec.~\ref{Sec:sphere}, we study spherical smectic and discuss two types of smectic states, i.e. hedgehog/latitudinal states and spiral states, on a sphere.  In Sec.~\ref{Sec:sphere-quasibb}, we use the formalism of holomorphic forms, which we develop in Appendix~\ref{Sec:holomorphic}, to address the third type of states for spherical smectics, i.e. quasi-baseball states.   Finally in Appendices \ref{Sec:holomorphic}, \ref{App:calculation}, \ref{App:Hodge-dual}, \ref{App:form-nematic}, we present some formalisms and calculation details that we used in the text.

\section{Analytic Description of Smectic Order on Curved Surface}
\label{Sec:Forms}

\subsection{Differential Forms and Density Wave Modulation} 
We shall only consider smectic ordering on two dimensional surfaces in this work.  The formalism, however, can be straightforwardly generalized to three dimensions.   A two dimensional curved surface, i.e. a Riemannian manifold, embedded in the three dimensional Euclidean space can be parameterized locally, using arbitrary curvilinear  coordinates $x = \{x^1,x^2\}$, as a vector function $ \rv(x^1,x^2) = \rv(x)$ which gives the position of the point $(x^1,x^2)$ in the embedding space.  A smectic order on the surface can be described by one dimensional density modulation Eq.~(\ref{density-0}).  We have already seen in Sec.~\ref{Sec:Intro} that the phase field $\Theta(x)$ is generically not globally defined as a real-valued function.  Nevertheless, if we choose a reference point $x$ with phase $\Theta(x)$, then the phase at a neighboring point $x + \delta x$ can be written as  
\begin{eqnarray}
\Theta(x+  \delta x) \approx  \Theta(x) + \psi_{\alpha}(x) \delta x^{\alpha}, 
\label{pssi-def}
\end{eqnarray}
as long as $\delta x$ is small comparing to all the other characteristic length scales.  Here $\{ \psi_{\alpha}, \alpha = 1,2\}$ is the local wave vector for the density wave around $x$, expressed in the generally non-orthogonal coordinate system $\delta x^{\alpha}$.

In modern differential geometry\footnote{For a physicists-friendly introduction of modern differential geometry, see reference \cite{book:Nakahara,book:Frankel,Stone-notes}. }, the object 
\begin{eqnarray}
\psi(x) = \psi_{\alpha} d x^{\alpha}
\label{psi-def}
\end{eqnarray}
is called {\em a differential form of rank one}, or {\em 1-form} in short.  The set $\{dx^1,dx^2\}$ constitutes a bases of the vector space of all 1-forms.   While the classical meanings of $dx^{\alpha}$ are infinitesimal displacements of coordinates, in modern differential geometry, they are defined as the dual vectors of the tangent vectors at point $x$, much like wave vectors are the duals to real space vectors.  Let $\{ \vec{e}_{\alpha} = \partial \rv/\partial x^{\alpha} \}$ be the natural basis vectors of the tangent plane at point $x$, an arbitrary tangent vector $\vec{V}$ can be expanded in terms of these as $\vec{V} = V^{\beta} \vec{e}_{\beta}$.   The differential forms $dx^{\alpha}$ act on a tangent vector $\vec{V}$ as linear operators and yield numbers:
\begin{eqnarray}
\langle dx^{\alpha}, V^{\beta} \vec{e}_{\beta} \rangle
= V^{\alpha},
\end{eqnarray}
i.e., they simply extract the $\alpha$-th components of the vector $\vec{V}$.  A general differential form Eq.~(\ref{psi-def}) then acts on the vector $\vec{V}$ as 
\begin{eqnarray}
\langle \psi, \vec{V} \rangle
= \psi_{\alpha} V^{\alpha}.
\end{eqnarray}

For a given 1-form $\psi$ at $x$, let us now consider all tangent vectors $\vec{V}$ at $x$ that satisfy $\langle \psi, \vec{V} \rangle = 0$.  It is clear that all these vectors form a one  dimensional subspace of the two dimensional tangent space.   We can think of this subspace as the infinitesimal version of a smectic layer that passes through $x$.  
Let us now consider a 1-form field defined on the manifold, i.e. there is a 1-form $\psi(x)$ defined at every point $x$.  We try to find a curve $\rv(s) = \rv(x^1(s), x^2(s))$ such that its tangent vector $\frac{d \rv}{ds}= \frac{d x^{\alpha}}{d s} \vec{e}_{\alpha}$ everywhere satisfies 
\begin{eqnarray}
\langle \psi,\frac{d \rv}{ds} \rangle  
= \psi_{\alpha} \frac{d x^{\alpha}}{d s} = 0. 
\label{psi-layer}
\end{eqnarray}
Such a curve can always be found in the neighborhood of $x$, by solving the above first order differential equation (which is usually nonlinear in $x^{\alpha}$).  Therefore a 1-form $\psi(x)$ locally defines a curve passing through an arbitrary given point $x$ via Eq.~(\ref{psi-layer}).

\begin{figure}
\begin{center}
\includegraphics[width=4cm]{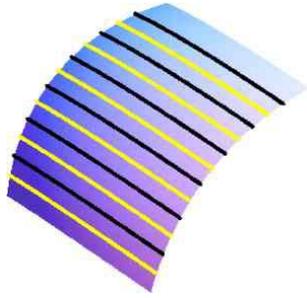}
\caption{(Color online) The state of black layers and that of pink layers differ only by a shift of constant phase, and therefore are described by the same differential form. }
\label{smectic-translation}
\end{center}
\vspace{-5mm}
\end{figure}

For an arbitrary smectic state on a curved substrate with no disclination, we can always choose a direction along which the phase field $\Theta$ increases.  The differential of the phase field $\psi = d\Theta$ is therefore always well defined, even though the phase field $\Theta$ itself may not be.  On the other hand, for a given 1-form field $\psi$, the phase at a point $y$ is related to that at point $x$ via
\begin{eqnarray}
\Theta(y) = \Theta(x) + \int_{x}^y \psi 
= \Theta(x) + \int_{x}^y \psi_{\alpha} dx^{\alpha}. 
\end{eqnarray}
For the moment we shall assume that this integral is independent of the path connecting $x$ with $y$.   We shall discuss what happens when this condition is violated below.   Hence we can pinpoint the exact location of all smectic layers, as soon as we know the phase at one single point $x$ and the 1-form field $\psi$. In other words, the phase field $\Theta$ is determined by the differential form $\psi$ only up to an unknown constant $\Theta(x)$.  Tuning of this constant is equivalent to translating all smectic layers in a common direction, as illustrated in Fig.~\ref{smectic-translation}, which can not be captured by the differential form $\psi$.  

1-forms therefore provide a natural local description of smectic ordering in arbitrary curved manifold.  This fact is actually well known in mathematics---mathematicians often visualize a 1-form field $\psi$ in three dimensional space as a set of onion skins (analogue of smectic layers!), and a tangent vector $\vec{V}$ as a needle with given length.  The pair $\langle \psi, \vec{V}\rangle$ simply measures the number of layers pierced through by the needle.

The metric tensor 
$$ \mm{g} = g_{\alpha\beta} \, dx^{\alpha} \otimes dx^{\beta}$$ 
is defined by the inner product between basis vectors:
\begin{eqnarray}
g_{\alpha\beta} = \partial_{\alpha} \rv \cdot \partial_{\beta} \rv
= \vec{e}_{\alpha} \cdot \vec{e}_{\beta}. 
\label{metric-def}
\end{eqnarray}
As usual, we can use the matrix $g_{\alpha\beta}$, as well as its inverse $g^{\alpha\beta}$ to raise and lower tensor indices.   A contravariant vector, i.e. tangent vector, $\vec{\psi}$ associated with the differential form $\psi$ can be defined as 
\begin{eqnarray}
\vec{\psi} = g^{\alpha\beta} \psi_{\alpha} \vec{e}_{\beta}
= \psi^{\beta} \vec{e}_{\beta}.  
\label{vec-psi}
\end{eqnarray}
We shall however use the same notation $\psi$ both for the 1-form and for the corresponding tangent vector $\vec{\psi}$ below.  This should not cause any confusion for a Riemannian manifold.  The norm, or magnitude, of the form $\psi$, is defined by 
\begin{eqnarray}
|\psi|^2 = |\vec{\psi}|^2 = g^{\alpha\beta} \psi_{\alpha} \psi_{\beta}
= g_{\alpha\beta} \psi^{\alpha} \psi^{\beta}.  
\label{layer-spacing}
\end{eqnarray}
The geometric meaning of $|\psi|/2\pi$ is {\em the number of smectic layers in unit distance along the layer normal. }  In other words, $2\pi/|\psi|$ is the normal distance between two neighboring smectic layers:
\begin{eqnarray}
d = \frac{2 \pi}{|\psi|}.  
\label{d-psi}
\end{eqnarray}  

In the purely geometric description of smectic orders, we do not distinguish the smectic layer normal and the nematic director field. Both are characterized by the same unit 1-form: 
\begin{eqnarray}
\nh \equiv \nh_{\alpha} dx^{\alpha} 
=\frac{1}{|\psi|} \psi = |\psi|^{-1} \psi_{\alpha} dx^{\alpha}. 
\end{eqnarray}
We shall also use the same notations for the corresponding unit tangent vector field
\begin{eqnarray}
\nh = \nh^{\alpha} \vec{e}_{\alpha}
= \frac{1}{|\psi|} \, \vec{\psi}  
= \frac{1}{|\psi|} \, \psi^{\alpha} \vec{e}_{\alpha}. 
\label{nh-psi-def}
\end{eqnarray}
Clearly 
\begin{eqnarray}
|\nh|^2 = g_{\alpha\beta} \nh^{\alpha} \nh^{\beta} 
= g^{\alpha\beta} \nh_{\alpha} \nh_{\beta}
= 1,
\end{eqnarray}
where $\nh^{\alpha} = g^{\alpha\beta}\nh_{\beta}$.

\subsection{Dislocations}
\label{sec:dislocations}
The exterior derivative, or antisymmetric derivative, of the 1-form $\psi$ is a differential form of rank two given by:
\begin{eqnarray}
\rho &=& d\psi = \partial_{\alpha}\psi_{\beta} \,dx^{\alpha} \wedge dx^{\beta}
\nonumber\\
&=& ( \partial_{1}\psi_{2} - \partial_{2}\psi_{1}) \,
dx^{1} \wedge dx^{2},
\label{dislocation-1}
\end{eqnarray}
where 
$$dx^{1} \wedge dx^{2}
 =  dx^{1} \otimes dx^{2}
 - dx^{2} \otimes dx^{1} 
 = - dx^{2} \wedge dx^{1}. $$

\begin{figure}
\begin{center}
\includegraphics[width=8cm]{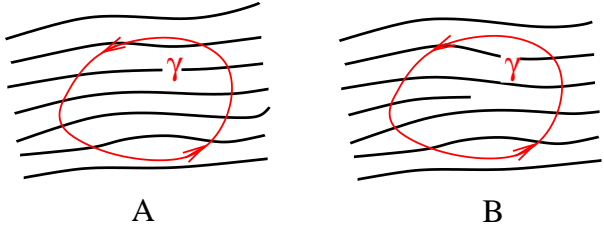}
\caption{A): The integral of $\psi$ along a close loop $\gamma$ vanishes if there
is no defects inside $\gamma$.  B): If there is a dislocation inside $\gamma$, the integral
is $2\pi$ times the dislocation charge.  }
\label{Burgers-circuit}
\end{center}
\vspace{-2mm}
\end{figure}

Consider a small loop $\gamma$ that can be continuously shrunk to a point~\footnote{Note that this would not be possible if there is hole inside $\gamma$. }, as shown in Fig.~\ref{Burgers-circuit}. According to the definition of the form $\psi$, the contour integral of $\psi$ along $\gamma$ is the total change of the phase field, $\Delta \Theta$, along $\gamma$.  Using Stoke's theorem we can transform this integral into a two dimensional integral over the enclosed region $D$:
\begin{eqnarray} 
\Delta \Theta =  \oint_{\gamma} \psi =  \int_{D} d\psi.  
\label{Delta-Theta}
\end{eqnarray}
If there is no defects inside $\gamma$, as shown in Fig.~\ref{Burgers-circuit}A, the overall change of the phase field $\Delta \Theta$ must be zero: 
\begin{equation}
d \psi = 0.
\end{equation} 
By contrast, if there is a dislocation inside the loop $\gamma$, as shown in Fig.~\ref{Burgers-circuit}B, the phase change  $\Delta \Theta$ along the loop must be the dislocation charge $n$ multiplied by $2\pi$:    
\begin{eqnarray}
\Delta \Theta = \oint_{\gamma} \psi = \int_D d\psi = 2 \pi n.   
\label{quantization}
\end{eqnarray}
Eq.~(\ref{quantization}) constitute the quantization condition that is to be satisfied by the differential form $\psi$, if it describes a smectic order.  The exterior differential $\rho = d\psi$ must be identified with the dislocation distribution.  The loop $\gamma$ and the integer $n$ are the analogues of Burgers' circuit and Burgers' vector in the case of crystalline dislocations.  If there are $k$ dislocations at places $\{x_1, \ldots,x_k\}$ with charges $\{n_1, \ldots,n_k\}$, the total dislocation density is given by
\begin{eqnarray}
&& \rho  = d\psi = 2 \pi \sum_{a=1}^k n_a\,
\delta^2(x-x_a) \, dx^1 \wedge dx^2,\\
&&\int d\psi = 2\pi \sum_{a = 1}^k n_a. 
\end{eqnarray}

In modern differential geometry, a differential form $\psi$ satisfying $d\psi = 0$ is said to be {\em closed}.  Hence {\em defects-free smectic states are described by closed 1-forms}.  
A 1-form $\psi$ is said to be {\em exact} if it can be written as the differential of a function $\Theta$: $\psi = d \Theta$.  An exact form is clearly closed, since
\begin{eqnarray}
 d \psi = d (d \Theta) 
= (\partial_{\alpha} \partial_{\beta} \Theta 
- \partial_{\beta} \partial_{\alpha} \Theta ) 
dx^{\alpha} \wedge dx^{\beta} = 0. 
\end{eqnarray}
This result is called the {\em Poincar\'e lemma}.  In the context of smectic ordering, it means that a smectic state where the phase field $\Theta(x)$ is globally well defined does not contain any the dislocation, a rather obvious result.

\subsection{Global Dislocations and Cohomology Theory}


In a flat space, if there is no dislocation everywhere, $d\psi = 0$, one can find a globally defined phase field $\Theta(x) $ such that $\psi = d\Theta$.  Consequently the integral Eq.~(\ref{Delta-Theta}) along arbitrary loop $\gamma$ always vanishes.   This is the inverse of the Poincar\'e lemma, which holds in flat space, but may fail in space with different topology.   One simple counter example on a cylinder is already shown Fig.~\ref{close-loop-cylinder} of Sec.~\ref{Sec:Intro}.   Here we give a more detailed discussion in terms of differential forms.   Using the cylindrical coordinate system ($z$, $\varphi$),  let us consider a smectic state
 \begin{eqnarray}
\psi = a \, dz + k\, d\varphi. \label{smectic-cylinder}
\end{eqnarray} 
One easily check that $d\psi = 0$, hence this state is dislocation-free. 
The metric tensor Eq.~(\ref{metric-def}) is given by 
\begin{eqnarray}
g_{\varphi\varphi} = R^2, \,\, 
g_{zz} = 1 \,\,
g_{\varphi z} = g_{z \varphi} = 0. 
\end{eqnarray}
The layer spacing can be calculated using Eq.~(\ref{layer-spacing}) and Eq.~(\ref{d-psi}):
\begin{eqnarray}
d = \frac{2\pi}{\sqrt{a^2 + R^{-2}k^2}},
\end{eqnarray}
which is independent of $z$ and $\varphi$.  Hence Eq.~(\ref{smectic-cylinder}) indeed describes a defects-free smectic state with constant layer spacing.  

The integral of $\psi$ over the loop $\gamma_{\varphi}$ wrapping around the cylinder once is given by
\begin{eqnarray}
\oint_{\gamma_{\varphi}} \psi = \int_0^{2\pi} k \,d \varphi
=  2 \pi \, k,  \label{smectic-cylinder-2}. 
\end{eqnarray}
Hence $k$ must be an integer; it is the {\em global dislocation charge} associated with the non-retractable loop $\gamma_{\varphi}$.  As discussed in Sec.~\ref{Sec:Intro}, $k$ is also the minimal number of smectic layers intersecting the non-retractable loop $\gamma_{\varphi}$.  
\footnote{Even though the form $\psi$ in Eq.~(\ref{smectic-cylinder}) looks like an exact form, it is really not.  This is because the angular coordinate $\varphi$ is not a single-valued function on the cylinder.  Hence there is not a single valued phase field $\Theta$ such that $\psi = d \Theta$.  The integral of an exact form $d \Theta$ over an arbitrary loop $\gamma$ vanishes identically, regardless of whether $\gamma$ is retractable or not. } 

The integral Eq.~(\ref{smectic-cylinder-2}) is independent of the shape and position of the loop $\gamma_{\varphi}$, as long as it winds around the cylinder once.   To see this, we note that the integer $k$ can only jump discretely.  However, in the absence of dislocation, such a jump can not happen as one continuously deforms the contour.  Likewise, one may continuously deform the smectic layers and the integer $k$ does not change.  The global dislocation charge $k$ therefore is invariant under arbitrary continuous deformation of the smectic state, i.e. it is a topological invariant.  It classifies all topologically distinct defects-free smectic states on a cylinder.

Global dislocations are consequences of the nontrivial topology of the substrate, and are naturally addressed by deRham's cohomology theory \cite{book:Nakahara,book:Frankel,Stone-notes}.   In a compact manifold, a non-retractable loop is called a 1-cycle, while a closed but not exact 1-form is called a 1-cocycle.  De Rham's theorem \footnote{The de Rham's theorem works for differential forms with arbitrary ranks.  What we invoke here is the special case of 1-forms. } of cohomology says that cycles and cocycles are dual to each other, and that there are equal number $b_1$ of independent cycles and independent cocycles.  This integer $b_1$ is called the first Betti number and is a topological property of the manifold.   Let $\{\beta_1, \beta_2, \ldots, \beta_{b_1}\}$ be the $b_1$ linearly independent cocycles, we can always find the $b_1$ cycles $\{\gamma_1,\gamma_2,\ldots,\gamma_{b_1}\}$ that form the dual basis, such that 
\begin{eqnarray}
\oint_{\gamma_i} \beta_j = 2 \pi \, \delta_{ij}.  
\end{eqnarray}
An arbitrary closed form $\psi$ can be expressed as a linear superposition of $\beta_i$ and an exact form $d \Phi$:
\begin{eqnarray}
\psi = N_1 \, \beta_1 + \ldots + N_{b_1} \,\beta_{b_1} +  d\Phi.  
\label{psi-expansion}
\end{eqnarray}
As will be discussed in detail in Sec.~\ref{Sec:heuristic}, the exact form $d\Phi$ describes elastic deformation, and therefore does not carry any topological content.  The closed form $\psi$ Eq.~(\ref{psi-expansion}) is exact if and only if all the coefficients $N_i$ are zero.   Furthermore, if $\psi$ describe a smectic order, it must satisfy the quantization condition Eq.~(\ref{quantization}), hence {\em all the coefficients $N_i$ in Eq.~(\ref{psi-expansion}) must be integers}.  They are precisely the global dislocation charges that classify all topologically distinct smectic states.  The first Betti number of cylinder is one; hence only one nontrivial cycle, i.e., the loop $\gamma_{\varphi}$ winding around the cylinder once, and only one nontrivial cocycle, i.e., $\psi = d \varphi $.  The 1-form Eq.~(\ref{smectic-cylinder})  is the linear superposition of this cocycle $d\varphi$ and an exact form $a\, dz$.

\subsection{Smectic Order on Torus}
\label{Sec:torus}

\begin{figure}
\begin{center}
\includegraphics[width=8cm]{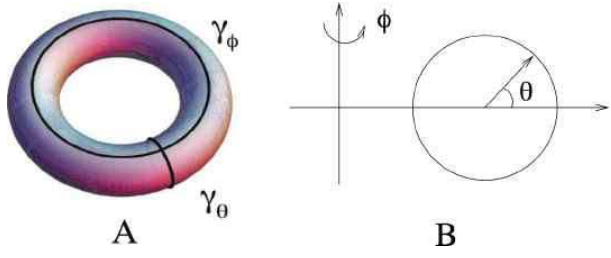}
\caption{A): Two non-retractabe loops on a torus. B) Coordinate system for torus.  
$ -\pi < \theta \leq \pi$, $0 \leq \phi < 2\pi$.}
\label{torus-cycles}
\end{center}
\vspace{-5mm}
\end{figure}

Let us apply the cohomology theory to defects-free smectic packing on a torus.   The first Betti number of a torus is $b_1 = 2$ \cite{book:Nakahara,book:Frankel}, hence there are two independent 1-cycles, $\gamma_{\theta}$ and $\gamma_{\phi}$, wrapping around the torus in two different directions, as shown in Fig.~\ref{torus-cycles}A.  Correspondingly there are two independent 1-cocycles, naturally chosen as $d\theta$ and $d\phi$, where $\theta$ and $\phi$ are the angular coordinates wrapping around the loops $\gamma_{\theta}$ and $\gamma_{\phi}$, see Fig.~\ref{torus-cycles}B.   We easily see that 
$\{\gamma_{\theta},\gamma_{\phi}\}$ is indeed the dual basis of $\{d\theta,d\phi\}$:
\begin{eqnarray}
&& \oint_{\gamma_{\theta}} d \theta = \oint_{\gamma_{\phi}} d\phi = 2 \pi,
\nonumber
\\
&& \oint_{\gamma_{\theta}} d \phi = \oint_{\gamma_{\phi}} d\theta = 0. 
\nonumber
\end{eqnarray}
$d\theta$ and $d\phi$ are not exact, since $\theta$ and $\phi$ are not single valued function.

\begin{figure}
\begin{center}
\includegraphics[width=12cm]{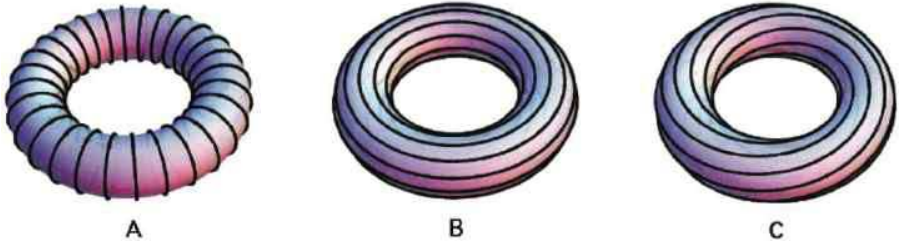}
\caption{Smectic packing state on torus with 
$(N_{\theta} = 0,N_{\phi } =28)$, $(N_{\theta} = 12,N_{\phi } = 0)$,
and $(N_{\theta} = 11,N_{\phi } = 5)$ respectively. The last one is left handed. }
\label{torus-states}
\end{center}
\vspace{-5mm}
\end{figure}

An arbitrary defects-free smectic state $\psi$ on a torus is therefore characterized two global dislocation charges $N_{\theta}$ and $N_{\phi}$: 
\begin{eqnarray}
\psi_{N_{\theta}, N_{\phi}} & = &
N_{\theta} \, d\theta + N_{\phi} \,d\phi
+ d\Phi(\theta,\phi)    
\label{packing-torus} 
\\
&= & ( N_{\theta} + \partial_{\theta} \Phi) \, d\theta 
+ ( N_{\phi} + \partial_{\phi} \Phi) \,d\phi, 
\nonumber
\end{eqnarray} 
where
\begin{eqnarray}
\oint_{\gamma_{\theta}} \psi = 2 \pi \, N_{\theta},\quad 
\oint_{\gamma_{\phi}} \psi = 2 \pi \, N_{\phi}. 
\nonumber
\end{eqnarray}
Two integers $(N_{\theta}, N_{\phi})$ classify all topologically distinct, defects-free smectic states on a torus, 
and are the minimal numbers of smectic layers intersecting the non-retractable loops $\gamma_{\theta}$ and $\gamma_{\theta}$ respectively.  Two states with different charges $N_{\theta}$ and $N_{\phi}$ are topologically different and therefore are mutually inaccessible by elastic deformations.  Three example states are shown in Fig.~\ref{torus-states}.  

\begin{figure} 
\begin{center}
\includegraphics[width=7cm]{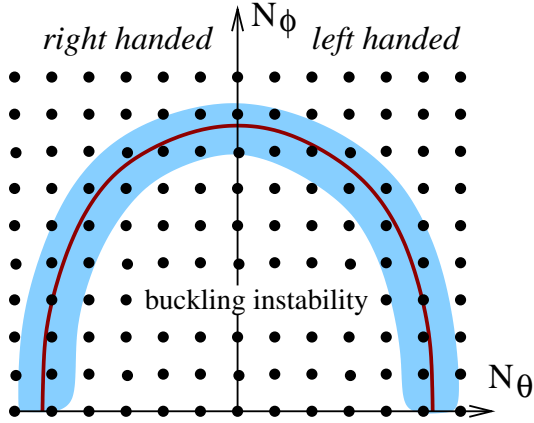}
\caption{Smectic states on a torus with distinct global dislocation charges are shown as grid points in the upper-half $(N_{\theta},N_{\phi})$ plane.  The curve (ellipse) is the locus of zero nonlinear strain $w = 0$, as determined by Eq.~(\ref{vanish-strain-locus}).  States in the shaded region have low strain energy and are approximately degenerate in the limit of large system size.  States well inside the ellipse are expected to exhibit buckling instability. } 
\label{strain-charges}
\end{center}
\vspace{-5mm}
\end{figure}

Since $\psi$ and $-\psi$ describe the same smectic state, according to Eq.~(\ref{packing-torus}), $(N_{\theta}, N_{\phi})$ and $(-N_{\theta}, -N_{\phi})$ correspond to the same topological class of states.   We shall therefore choose the convention $N_{\phi} \geq 0$ to avoid double counting. Hence a defects-free state Eq.~(\ref{packing-torus}) can be represented as a point in the lattice in the upper half $(N_{\theta},N_{\phi})$ plane, as shown in Fig.~\ref{strain-charges}.  On the other hand, under spatial inversion of the three dimensional embedding space,  the coordinates $(\theta,\phi)$ transform as 
\begin{eqnarray}
\theta \rightarrow - \theta, \quad \phi \rightarrow \phi + \pi. 
\end{eqnarray}
The 1-form Eq.~(\ref{packing-torus}) then transforms as
\begin{eqnarray}
{\mathcal I} \,\, \psi_{N_{\theta}, N_{\phi}} = 
- N_{\theta} \, d\theta + N_{\phi} \,d\phi
+ d\Phi(- \theta,\phi+ \pi).  
\end{eqnarray}
Since translation of $\phi$ by $\pi$ amounts to rotation of the torus by $\pi$ around the $z$ axis, we find that a state $(N_{\theta}, N_{\phi})$ is transformed into $(- N_{\theta}, N_{\phi})$ by the space inversion.   Consequently states with $N_{\theta} \neq 0$ and $N_{\phi} \neq 0$ are not invariant under spatial inversion, i.e. they are chiral.   We shall call states with $N_{\theta}>0$ left handed, while states with $N_{\theta}<0$ right handed.  Clearly a left handed state is transformed into a right hand state under spatial inversion, and vice versa.

To understand the geometric properties of the states Eq.~(\ref{packing-torus}) let us 
calculate the layer spacing for the special case with $\Phi = 0$.  With the coordinate system shown in Fig.~\ref{torus-cycles}B, the torus is parameterized by two angles $(\theta,\phi)$ as 
\begin{eqnarray}
\rv(\theta,\phi) = \left( 
\begin{array}{c}
(R_{\phi} + R_{\theta} \cos\theta) \cos \phi \\
(R_{\phi} + R_{\theta} \cos\theta) \sin \phi \\
R_{\theta} \sin\theta
\end{array}
\right). 
\end{eqnarray}
The corresponding metric tensor can be easily calculated using Eq.~(\ref{metric-def}):
\begin{eqnarray}
&& g_{\theta\theta} = R_{\theta}^2, \nonumber\\
&&g_{\phi \phi} = (R_{\phi} + R_{\theta} \cos \theta)^2,
\nonumber\\
&& g_{\theta \phi} = g_{\phi \theta} = 0. 
\end{eqnarray}
We shall also define the aspect ratio
\begin{equation}
r = \frac{R_{\theta}}{R_{\phi}}.  
\label{aspect-ratio-def} 
\end{equation}
Note that $0<r<1$ if the cylinder does not self-intersect.  We shall be especially interested in the thin torus limit $r \ll 1$, where $r$ provides a natural small parameter that greatly simplifies the analysis.

Defining two parameters $n_{\theta}$ and $n_{\phi}$ as follows
\begin{eqnarray}
n_{\theta} \equiv \frac{1}{2 \pi R_{\theta}} 
 N_{\theta},  \quad
n_{\phi} \equiv \frac{1}{2 \pi R_{\phi}} 
N_{\phi},  
\label{n-def}
\end{eqnarray}
the layer spacing of the state Eq.~(\ref{packing-torus}) can then be calculated using Eq.~(\ref{d-psi}):
\begin{eqnarray}
d^{-2} &=&  \frac{|\psi|^2}{4 \pi^2} = 
\left( n_{\theta}^2 + \frac{n_{\phi}^2}{(1+ r \cos \theta)^2} \right)
\label{torus-spacing}\\
&=&  \left( n_{\theta}^2 +  n_{\phi}^2 \right)
\left( 1 - \frac{2 r \cos \theta}{1 +  n_{\theta}^2/n_{\phi}^2}
+ O(r^2)
\right), \nonumber
\end{eqnarray}
where we have expanded in terms of the aspect ratio $r$ up to the linear order.  To the zero-th order therefore the layer spacing is independent of coordinates $(\theta,\phi)$.  Hence in the thin torus limit $r \rightarrow 0$, the layer spacing is strictly constant for arbitrary charges ($N_{\theta}, N_{\phi})$.     
For nonzero aspect ratio $r$, the variation of layer spacing is relatively small if either of the following conditions is satisfied:
\begin{eqnarray}
r \ll1 \quad \mbox{or} \quad  
{n_{\theta}^2}\gg {n_{\phi}^2} . 
\end{eqnarray}
These are the cases where we expect that the defects-free smectic states Eq.~(\ref{packing-torus}) have low free-energy.   By contrast, on a fat torus and if ${n_{\theta}^2} \ll {n_{\phi}^2}$, 
the state Eq.~(\ref{packing-torus}) has significant variation in layer spacing, and therefore costs large amount of strain energy.  Dislocations are expected to proliferate to reduces the elastic strain energy.   More detailed study of the energetic issues for torroidal smectic will be presented in Sec.~\ref{Sec:energetics}.

Let us determine the loci of smectic layers for a state Eq.~(\ref{packing-torus}) with
$\Phi  =0$.   Without loss of generality, let us assume $N_{\theta} \neq 0$.  Let us choose the coordinate system $(\theta,\phi)$ such that one layer, parameterized by $(\theta(t), \phi(t))$, passes through point $\theta = \phi = 0$.   This layer then must satisfy the equation:
\begin{eqnarray}
N_{\theta}\frac{d \theta}{dt} + N_{\phi} \frac{d\phi}{dt} = 0,
\end{eqnarray} 
whose solution is simply given by 
\begin{eqnarray}
N _{\theta} \,\theta + N_{\phi} \phi = 0.  
\end{eqnarray}
If one follows this layer and wraps around the torus in $\theta$ direction once, one obtains $\Delta \Theta = N_{\theta} \times 2 \pi$, and therefore ends up with $N_{\theta}$-th layer.  
The $N_{\theta} - 1$ layers between 0-th layer and $N_{\theta}$-th layer1 are described by the set of equations 
\begin{eqnarray}
N _{\theta} \,\theta + N_{\phi} \phi = 2 k \pi,
\quad k =  1,2, \ldots, N_{\theta} - 1.  
\end{eqnarray}
This is in fact how the states in Fig.~\ref{torus-states} are generated. 

The formalism of differential forms also allows us to deduce interesting properties of smectic states with dislocations on a torus.  For example, let $\psi$ describe a smectic state on a torus.  The total dislocation charge is then given by the integral of the dislocation density $d\psi$ over the whole manifold: 
\begin{eqnarray}
N_{\rm tot} = \int_{T^2} d \psi = \oint_{\partial  T^2} \psi = 0.
\label{partial-T}
\end{eqnarray} 
This is because a compact manifold such as $T^2$ by definition does not have a boundary. Hence the right hand side of Eq.~(\ref{partial-T}) must vanish.   Therefore the total dislocation charge of smectic on a compact surface must be zero.  Obviously, the same result holds for crystalline order on arbitrary compact manifold as well.  

\section{What do we mean by topological properties?}
\label{Sec:heuristic}
The precise definition of topology as a mathematical concept is quite involved.  Loosely speaking, however, {\em a topological property is invariant under continuous deformations.}  Mathematicians are primarily interested in the topological property of manifolds themselves, hence they consider continuous deformation of the manifolds.   Physicists, however, are usually interested in the topological properties of field configurations defined on manifolds, the relevant continuous deformations are therefore those of the fields on a fixed manifold.  This distinction is seldom made clear in the literature.  Of course, topology of fields is intrinsically related to that of the manifold where it is defined; by studying one we can learn about the other.  This relation is indeed the origin of the global dislocation charges discussed in this work.  


We are therefore interested in the properties of low energy smectic states on curved surfaces that are invariant under arbitrary continuous deformations.  Using the purely geometric description, a smectic state described as a stack of infinitesimally thin layers.   By a continuous deformation we mean a continuous distortion of the layers such that none of them gets broken; such a deformation shall also be called an {\em elastic deformation} for obvious reasons.  We shall define two smectic states to be topologically identical, or equivalent, to each other, if there exists an elastic deformation that brings one state to the other.   By definition elastic deformations do not change any topological property.  By contrast, any deformation that breaks some layers shall be called {\em plastic}, and  necessarily change some topological properties.  The concepts of elastic as well as plastic deformation obviously apply to all systems with translational orders, and reduce to the conventional definitions in the case of crystals.

\begin{figure}
\begin{center}
\includegraphics[width=8cm]{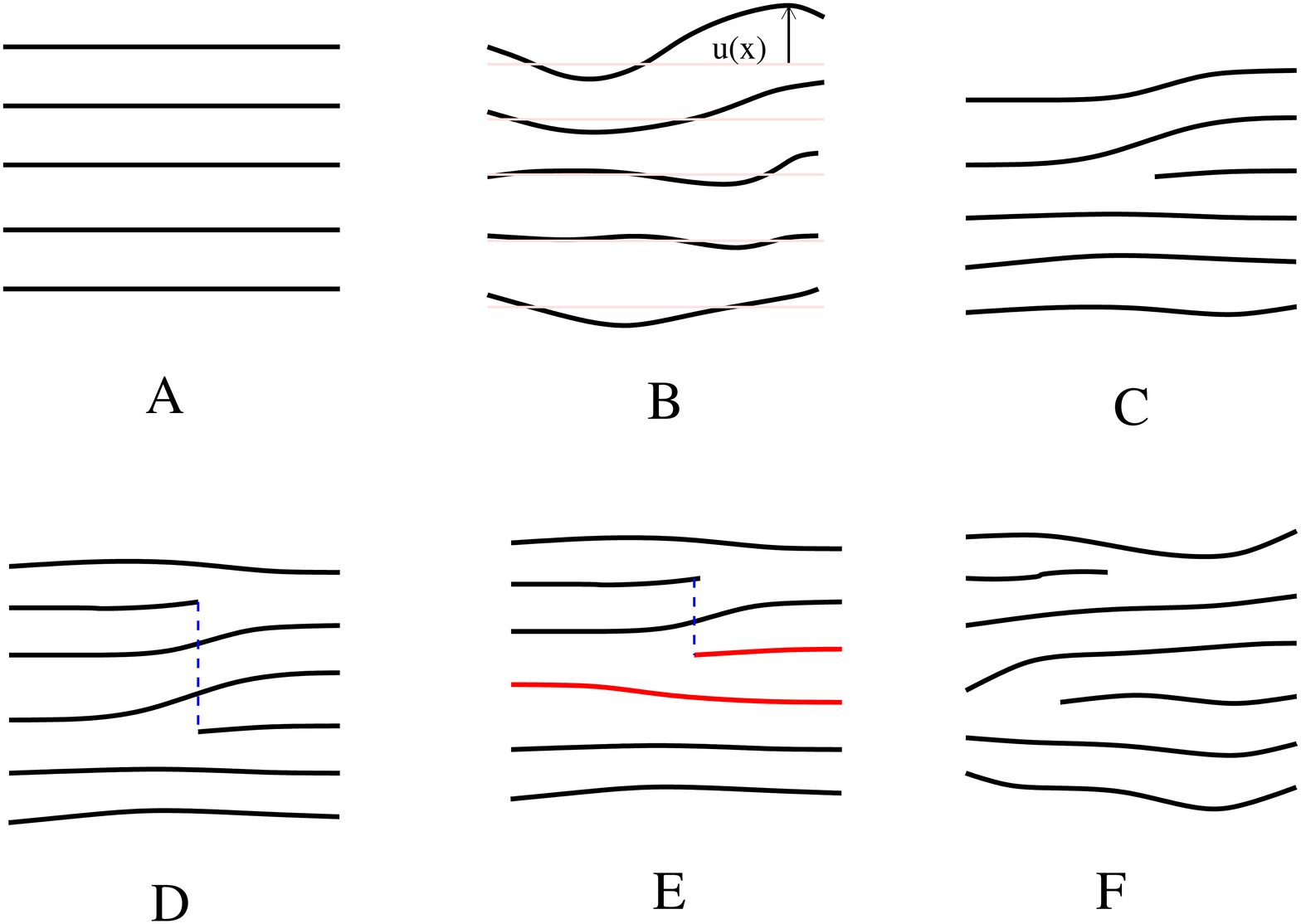}
\caption{Topology of smectic in flat space. A): ground state; B) defects-free deformed state that is topologically identical to A); C) a state with isolated dislocation with charge one; D) two dislocation with internal distance two; E) two dislocations with internal distance one; F) two dislocation with internal distance two. The deformation brings D) to E) is plastic; the one brings D) to F) is plastic.  States D) and F) are topologically identical.}
\label{smectics-states}
\end{center}
\vspace{-5mm}
\end{figure}

A number of results follow directly from the above definitions.   Layer spacing is not a topological quantity; neither is a kink, i.e. slope discontinuity, of layers.   In a flat space with no boundary, all defects-free states (Fig.~\ref{smectics-states}B) are topologically identical to the perfectly ordered ground state (Fig.~\ref{smectics-states}A) where all layers are flat and equal-spaced.  Any state with a single dislocation with charge $k$ (Fig.~\ref{smectics-states}C) is clearly topologically distinct from the ground state.   Two such states with distinct charges are also topologically distinct from each other.   The dislocation charge is therefore a topological property.   For states with two dislocations (Fig.~\ref{smectics-states}D, E, F), the number of layers between them is a topological property.   This number is the ``internal'' distance \footnote{We adopt this terminology from Kroner's study \cite{Kroner-notes} of crystalline defects using differential geometry. }, i.e. the distance measured by the local smectic layer spacing, between them.   To change this internal distance (Fig.~\ref{smectics-states}D to E), it is necessary to break and reconnect smectic layers.  In contrast, the physical distance between dislocations can be changed by changing layer spacing and without change of the internal distance (from Fig.~\ref{smectics-states} D to F); such a deformation is elastic.   For states with many dislocations, the internal distance between two dislocations depends on the path connecting them. 
 
A plastic deformation that breaks infinite number of layers shall be called a {\em non-local plastic deformation}.   For example, one only needs to cut three layers in order to go from the defects free state Fig.~\ref{smectics-states}A to the state Fig.~\ref{smectics-states}D with a pair of opposite dislocation pair.  Such a process is local, and can spontaneously happen in thermal equilibrium.   On the other hand, it takes a nonlocal plastic deformation to go from Fig.~\ref{smectics-states}A to Fig.~\ref{smectics-states}C: one needs to cut infinite number of layers in order to introduce an isolated dislocation into a defects free state.   



The homotopy theory of defects \cite{Homotopy-review-Mermin,Homotopy-review-Michel,Homotopy-review-Trebin,Homotopy-review-Kleman}, developed in 1970's, deals with deals with defects in various ordered media in Euclidean spaces.   In this theory, a topological property is defined to be invariant under arbitrary local deformations.   For systems with translational orders, such as smectic or crystals, these local deformations include both elastic and plastic deformations.  \footnote{For system with orientational orders, the distinction between plastic deformation and elastic deformation is irrelevant. }  For example, the states in Fig.~\ref{smectics-states}A\&D are topologically equivalent according to the homotopy theory.  {\em Topological properties according to our definition are therefore not necessarily topological according to the homotopy theory. }   

As stressed by Mermin~\cite{Homotopy-review-Mermin} long ago, the homotopy theory of topological defects, though rigorous and elegent, yields lots of strange results when applied to translational orders.  For example, by transporting a dislocation around a disclination, the Burgers vector of the former can be changed.  Consequently, dislocation almost completely loses its topological identity in the presence of disclinations.  The Homotopy theory therefore is not able to distinguish states with same disclination charge, but otherwise are widely different.  While these results are not A priori wrong, they are very counter-intuitive and often misleading.   It is therefore fair to say that the homotopy theory is not refined enough to capture the differences and connections 
between orientational defects and the translational ones in a translationally ordered system.    This point may become clearer when we discuss smectic order on sphere in Sec.~\ref{Sec:sphere}\&\ref{Sec:sphere-quasibb}.  


Differential forms, augmented by de Rham's cohomology theory, as discussed in this work, provides a natural frameworks for study of translational orders both in flat space and in curved space.   The topological significance of defects pairs, as shown in Fig.~\ref{smectics-states}D, E \&F, is captured by differential forms that are not closed; elastic deformations are naturally described by exact differential forms.   Being a geometric description, differential forms also provide a natural description for the energetics of translational ordered systems.  The issue of disclinations is more subtle, and in principle should be described by nontrivial fiber bundle structure associated with the differential forms.   For  two dimensional smectics, however, there is an easy alternative in terms of multiple valued complex differential forms, as will be discussed in Sec.~\ref{Sec:sphere-quasibb}.  A systematic exploration of disclinations in translational systems in terms of fiber bundle theory shall be left to a future work.  Finally we note that this geometry-based approach is in the same vein as the intuitive Volterra treatment of dislocation and disclination in continuum solids.   


\subsection{Elastic Deformations as Exact Differential Forms} 

It is a well known result in differential geometry that the integral of a closed 1-form along a loop is a topological quantity, that is, it is invariant under arbitrary continuous deformation of the loop and of the differential form itself.  This is the mathematical origin of the global dislocation charges that we discuss in the preceding section.~\footnote{Similar phenomena also arise in lattice gauge theory~\cite{LGT-review-Kogut} with periodic boundary conditions.  The relevant space is then the $d$-dimensional torus, with $d$ independent non-retractable loops.  The integral of the gauge potential (a 1-form) along these loops then give $d$ topological charges that are invariant under arbitrary gauge transformations.  Hence the configuration space consists of many components that are not mutual accessible by any local operation.}  In this section, we study in detail how elastic deformations change the differential form describing the smectic order.  

\begin{figure} 
\begin{center}
\includegraphics[width=4cm]{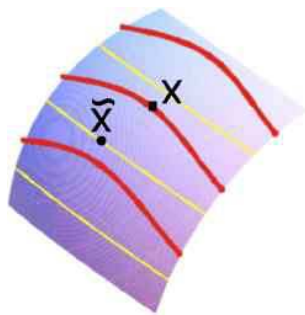}
\caption{(Color online) Yellow lines are undeformed smectic layers. Red lines are deformed layers.
The point (dot) with coordinates $\tilde{x}$ is deformed to the point (square) with coordinates $x$.  $x$ and $\tilde{x}$ are therefore the Eulerian as well as Lagrangian coordinates for the deformed point (square).  }
\label{smectic-deformation}
\end{center}
\vspace{-5mm}
\end{figure}

Let us consider a defects-free smectic state on a torus, described by a differential form Eq.~(\ref{packing-torus}).  It is characterized by two global dislocation charges $(N_{\theta},N_{\phi})$.   Let us elastically deform the substrate, together with the smectic pattern, such that a point with $(\tilde{\theta},\tilde{\phi})$ is brought to $(\theta,\phi)$ by the deformation, as illustrated schematically in Fig.~\ref{smectic-deformation}.   Using the convention of elasticity theory, we shall call $(\theta,\phi)$ the Eulerian coordinates and $(\tilde{\theta},\tilde{\phi})$ the Lagrangian coordinates.  The elastic deformation then define $(\tilde{\theta},\tilde{\phi})$ as functions of $(\theta,\phi)$ and vice versa.   The elastically deformed smectic state can then be obtained by simply replacing in Eq.~(\ref{packing-torus}) the coordinates $(\theta,\phi)$ by the Lagrangian coordinates  $(\tilde{\theta},\tilde{\phi})$, which shall in turn be treated as function of the Eulerian coordinates $(\theta,\phi)$:
\begin{eqnarray}
\psi_{\rm def} = 
N_{\theta} \, d \tilde{\theta}(\theta,\phi) + N_{\phi} \,d\tilde{\phi}(\theta,\phi)
+ d\Phi (\tilde{\theta},\tilde{\phi})   
\end{eqnarray}
Let us now look at the difference $\Delta \psi$ between $\psi_{\rm def}$ and $\psi$:
\begin{eqnarray}
&& \Delta \psi = \psi_{\rm def} - \psi
\nonumber\\
&=& N_{\theta} (d \tilde{\theta} - d \theta)
 + N_{\phi} (d \tilde{\phi} - d \phi)
\nonumber\\
  &+& d \left( \Phi(\tilde{\theta}, \tilde{\phi}) - \Phi(\theta, \phi)
  \right) .  
\end{eqnarray}
Because the Eulerian coordinates and the Langrangian coordinates are related by elastic deformation, the angles $\theta$ ($\phi$) wraps around the torus if and only if $\tilde{\theta}$ ($\tilde{\phi}$) does the same.   This implies that the integrals of $\Delta \psi$ both along the non-retractable loop $\gamma_{\theta}$ and along $\gamma_{\phi}$ vanish.   Hence, according to the de Rham's theorem, the integral of $\Delta \psi$ along arbitrary loop vanish and $\Delta \psi$ is an exact differential form.  Therefore the effect of an elastic deformation is to add an exact 1-form $\Delta \psi$ to the original differential form $\psi$, and does not change either global dislocation charge.  

An infinitesmial elastic deformation can be generated by a vector field $\vec{X} = X^{\alpha} \vec{e}_{\alpha}$.  Here for convenience we use the shorthand $x^{1} = \theta, x^{2} = \phi$.   Mathematically this means that a point $\tilde{x}$ (Lagrangian coordinates) is deformed to be the point $x$ (Eulerian coordinates) such that
\begin{eqnarray}
x^{\alpha} = \tilde{x}^{\alpha} + \epsilon \, X^{\alpha},
\end{eqnarray}
where $\epsilon$ is the infinitesimal flow parameter, and $X^{\alpha}$ are considered as functions of the Lagrangian coordinates $\tilde{x}^{\beta}$.   The deformed 1-form is then given by 
\begin{eqnarray}
\psi_{\rm def} &=& \psi_{\alpha} (\tilde{x}) d\tilde{x}^{\alpha} 
= \psi_{\alpha} (x - \epsilon X) d(x^{\alpha} - \epsilon \, X^{\alpha})
\nonumber\\
&=&  \psi_{\alpha}(x) dx^{\alpha} 
- \epsilon \left( 
\psi_{\alpha} \partial_{\beta} X^{\alpha}
+ X^{\alpha} \partial_{\beta} \psi_{\alpha}
\right) dx^{\beta},
\end{eqnarray}
where we have only kept terms up to linear order in $\epsilon$.   We recognize that the coefficient of $\epsilon$ is negative of the Lie derivative of the 1-form $\psi$.  Hence the infinitesimal change of $\psi$ induced by the elastic deformation is given by 
\begin{eqnarray}
\Delta \psi &=& \psi_{\rm def} - \psi = - \epsilon \, {\mathcal L}_{\vec{X}} \psi.
\label{delta-psi}
\end{eqnarray}
Using the infinitesimal homotopy relation~\cite{book:Nakahara,book:Frankel,Stone-notes}
\begin{eqnarray}
{\mathcal L}_{\vec{X}} = d \, i_{\vec{X}} + i_{\vec{X}} \,d, 
\end{eqnarray}
where $i_{\vec{X}}$ is the interior product, as well as the fact that $d\psi = 0$,  Eq.~(\ref{delta-psi}) reduces to  
\begin{eqnarray}
\Delta \psi  =   d \, \left( - \epsilon \, i_{\vec{X}} \psi \right)
= d \, \left( - \epsilon \, \psi_{\alpha} X^{\alpha} \right).
\label{dela-psi-2}
\end{eqnarray}
We have therefore explicitly constructed the potential corresponding to the exact form $\Delta \psi$ that arises from an elastic deformation generated by the vector field $\vec{X}$.  

Conversely, an arbitrary change of the 1-form $\psi$ by an exact form $d \Phi$ can be generated by an elastic deformation.   This elastic deformation is however not unique. 
In the case of torus, for example we can write Eq.~(\ref{packing-torus}) as
\begin{eqnarray}
\psi &=& N_{\theta} d\left( 
\theta + \frac{1}{N_{\theta}} \Phi(\theta,\phi)
 \right)  + N_{\phi} d \phi
 \nonumber\\
&=& N_{\theta} \,d\tilde{\theta}  + N_{\phi} \, d \tilde{\phi}, 
\end{eqnarray}
where
\begin{eqnarray}
\tilde{\theta} = \theta + \frac{1}{N_{\theta} } \Phi(\theta,\phi), \quad
\tilde{\phi} = \phi. 
\end{eqnarray}
According to our preceding discussion, then, this smetic state $\psi$ can be generated from the prototypical state 
\begin{eqnarray}
\psi_{\rm orignal} = N_{\theta} \,d\theta + N_{\phi} \, d\phi
\end{eqnarray}
 by the elastic deformation 
\begin{eqnarray}
(\tilde{\theta}, \tilde{\phi}) \longrightarrow (\theta, \phi), 
\end{eqnarray}
Again $(\theta,\phi)$ are the Eulerian coordinates and while $(\tilde{\theta},\tilde{\phi})$ are the Lagrangian coordinates.   This result provides a starting point for systematic studies of thermal elastic fluctuations in defects-free, translationally ordered system on a torus.

\section{Geometry and Energetics of Smectic on Curved Substrates}
\label{Sec:energetics}
Our problem is to study the topological as well as geometric properties of the low energy smectic states on a {\em rigid} and compact substrate,    This mission can be decomposed into two steps.  Firstly we use de Rham's theorem to classify all low energy smectic states with minimal number of defects.  For the case of torus, for example, this is achieved by Eq~(\ref{packing-torus}).  Note that the first step analysis is independent of energetics. As the second step, we minimize the total free energy, for given values of the global dislocation charges $N_{i}$, over the exact form $d\Phi$.  Physically, this means we let the system elastically relaxed to the energy minimum in a given topological sector.    A similar two-step process is also taken in the linear theory of topological defects in flat space, for example, in smectic and in crystals \cite{CMP:CL,CMP:Nelson}, where one first finds a singular part of the phonon field that gives the right topological charge, then adds a smooth part to this phonon field to obtain the energy-minimizing solution.   The same strategy is also implicitly employed in the Volterra's intuitive construction of defects in solids.     

To carry out the second step, however, we do need to understand the energetics of smectic on an arbitrary rigid substrate.  The total elastic free energy of a smectic consists of two parts.  Firstly deviation of the layer spacing away from its energetically preferred value $d_0$ costs elastic free energy.  The layer compression/dialation can  be captured by a dimensionless nonlinear strain $w$ defined as 
\begin{eqnarray}
w(x) =   \left( 1- \frac{d_0^2}{d^2}  \right)
 = \left( 1- \frac{d_0^2}{4\pi^2} |\psi|^2  \right), 
\label{strain-def}
\end{eqnarray}
where used was Eq.~(\ref{d-psi}).   The sign of $w$ is chosen according to the convention in elasticity theory, i.e. $w > 0$ ($w < 0$) corresponds to a dilational (compressional) strain, where the layer spacing increases (decreases).  The corresponding strain free energy is 
\begin{eqnarray}
F_B = \frac{B}{2} \int d \omega \,w(x)^2,
\label{F-B}
\end{eqnarray}
where $d\omega = \sqrt{g} \, dx^1\wedge dx^2$ is the invariant volume form.  It can be easily shown that this strain is identical to the one used in reference \cite{santangelo:017801},  and also reduces to the well known nonlinear strain for smectic order in flat space \cite{LC:deGennes,CMP:CL}.  

Secondly, variation in the layer normal, or bending of the layers, also costs elastic free energy.  The elastic free energy of vector/nematic/hexatic orders constrained to the tangent space of a flexible membrane has been explored extensively by many authors \cite{David-hexatic,PhysRevLett.67.1169,Helfrich-Prost-PRA-88,Nelson-Powers-93,Frank-Kardar-PRE-08}.  The most general form of this elastic free energy contains many independent parameters.  Here we shall focus on the simpler case of smectic orders on a rigid substrate, and leave the more complicated case of flexible substrate for a future work.   We first emphasize that it is {\em the variation of the layer normal in the three dimensional physical space} that costs elastic free energy.  In general, the elastic free energy depends on the director field $\nh$ as well as its covariant derivatives $D_{\alpha} \nh$.  It can also depend on the metric tensor $g_{\alpha\beta}$, 
the antisymmetric tensor $\epsilon^{\alpha \beta} = \sqrt{g^{-1}} \, \gamma^{\alpha\beta}$, with $\gamma^{\alpha \beta}$ the antisymmetric symbol, as well as the 
the extrinsic curvature tensor $\mathbf{K}$ with components $K_{\alpha\beta} = \hat{N} \cdot D_{\alpha} \vec{e}_{\beta} $, where 
\begin{eqnarray}
\hat{N} = \frac{\vec{e}_1 \times \vec{e}_2}{|\vec{e}_1 \times \vec{e}_2|},
\end{eqnarray}
is the unit normal vector of the surface.    In the present case of a rigid substrate, however, these latter objects are just fixed parameters, not dynamic quantities.  All tensor indices must be properly contracted over to ensure the reparameterization invariance.  Furthermore, the $\pi$ rotation symmetry pertaining to the nematic order requires that the elastic free energy must be invariant under the transformation $\nh \rightarrow -\nh$.   This is in strong contrast with the tilt order in double layers extensively studied long time ago~\cite{Nelson-Powers-93}.   
Furthermore, we shall consider only surfaces whose two sides are equivalent, i.e., those invariant under  $\hat{N} \rightarrow - \hat{N}$, so that terms like $\nh \cdot \mathbf{K} \cdot \nh$ are not allowed in the free energy.  


Incorporating all these constraints we find that there are six terms in total which are quadratic in $\nh$ and are marginally relevant under naive power counting~\footnote{We note that these terms can also be expressed in terms the unit tangent vector $\hat{t}$ of the smectic layers, where $\hat{t} \cdot \nh = 0$.  }.  First of all there are two terms involving the covariant derivatives of the director field: 
\begin{eqnarray}
F^0_{\nh} =  \int d\omega \left[ \frac{K_1}{2}
 (D \cdot \nh)^2 
+ \frac{K_3}{2 } 
(D \times \nh)^2 
\right],
\label{Frank-energy}
\end{eqnarray}
where $K_1$ and $K_3$ are splay and (inplane) bending Frank constants.  Note that 
\begin{eqnarray}
(D \times \nh) = \epsilon^{\alpha \beta} D_{\alpha} \nh_{\beta}
\end{eqnarray}
is a pseudo-scalar in two dimensions.   These two terms can be obtained by naively promoting the 2D flat space Frank free energy for nematics to curved space straightforwardly.  In the absence of dislocation, however, the inclusion of bending term with coefficient $K_3$ is not essential, since as is well known (also see discussion in Sec.~\ref{Sec:sphere}), bending of nematic director necessarily leads to change of smectic layer spacing, which is already penalized by the strain energy Eq.~(\ref{F-B})~\cite{Santangelo-Kamien-2005}.

Secondly there are three independent terms quadratic in the extrinsic curvature tensor 
$K_{\alpha\beta}$ \cite{Helfrich-Prost-PRA-88,David-hexatic,Nelson-Powers-93,Frank-Kardar-PRE-08}: 
\begin{eqnarray}
\frac{J_1}{2} (\nh \cdot \mathbf{K} \cdot \nh)^2 
+ J_2 (\nh \cdot \mathbf{K} \cdot \hat{t} )^2
+ \frac{J_3}{2}( \hat{t} \cdot \mathbf{K}  \cdot \hat{t} )^2,
\label{three-terms}
\end{eqnarray}
where $\hat{t}$ is the unit vector tangent to the smectic layers, i.e. perpendicular to $\nh$,
with $\hat{t}_{\alpha} = \epsilon_{\alpha}^{\,\,\beta} \nh_{\beta}$, while raising and lowering of indices using the metric tensor is implicit in the above expressions.    It is interesting to note that in the special case of a sphere, the extrinsic curvature tensor is isotropic, hence all these three terms reduce to trivial constants and therefore can be safely ignored. 

Finally for chiral systems, a term linear in $\epsilon_{\alpha\beta}$ is also allowed:
\begin{eqnarray}
c \, \nh \cdot \epsilon \cdot \mathbf{K}  \cdot \nh
= c \, \hat{t} \cdot \mathbf{K}  \cdot \nh.  
\label{chiral-term}
\end{eqnarray} 
We shall, however, not consider the effect of chiralty in this work.  

All the above terms were explicitly listed in reference \cite{Nelson-Powers-93} for the case of symmetric bi-layer with tilt order.   Also discussed there are some other terms that are allowed for tilt order, but not for nematic order.   For hexatic membranes, none of the terms in Eq.~(\ref{three-terms}) or in Eq.~(\ref{chiral-term}) is allowed.  Furthermore, the sixfold rotation symmetry of hexatic order requires $K_1$ and $K_3$  in Eq.~(\ref{Frank-energy}) to be the same.

Geometrically, the terms in Eq.~(\ref{three-terms}) express nothing but the energy cost for the director bending due to the extrinsic curvature of the substrate.  There is no symmetry principle that allows us to determine the relative importance of coefficients of the above terms~\footnote{The elastic model considered in \cite{santangelo:017801} corresponds to $K_1 = J_3 = K$ and all other moduli vanish.  A more general form of elastic model is considered in \cite{Frank-Kardar-PRE-08}. }.   Nevertheless, thermodynamic stability does require both $K_1$ and $K_3$ in Eq.~(\ref{Frank-energy}) to be positive.   Otherwise, the system would develop in-plane nematic director modulation spontaneously.  The coefficients $J_i$ in Eq.~(\ref{three-terms}) are more subtle: if $J_1 = J_2 = J_3 = J$, we easily see that Eq.~(\ref{three-terms}) reduces to $\frac{1}{2} J \,  \Tr \, \mathbf{K} \cdot \mathbf{K},$
which is independent of the dynamic variables $\nh$ and $\hat{t}$.  For a fluctuating liquid crystalline film, this term simply renormalizes the elastic modulus of extrinsic curvature.  By contrast, for a rigid substrate, which is our primary interest in this work, this term is just an irrelevant constant that can be safely ignored.   For the latter case, therefore, only the differences between $J_i$'s have physical meanings and the signs of the coefficients $J_i$ can not be determined by any physical consideration.   

Below we shall consider a particularly simple limit where $J_1 =2\, J_2 =  K_1 = K_3 = K$ and $J_3 = 0$.  Simple calculation then shows that Eq.~(\ref{three-terms}) reduces to $K/2 \, \nh  \cdot \mathbf{K}^2 \cdot \nh$, and the sum of Eq.~(\ref{Frank-energy})  and  Eq.~(\ref{three-terms}) reduces to: 
\begin{eqnarray}
F_{\nh} = \frac{K}{2} \int d\omega\, g^{\alpha \beta} \partial_{\alpha} \nh \cdot
\partial_{\beta} \nh,
\label{F-n}
\end{eqnarray}
where $\partial_{\alpha}$ are the {\em ordinary derivatives}, while $\nh$ is constrained to be  tangential to the substrate.   We note that Eq.(\ref{F-n}) is the elastic free energy for a {\it nonlinear Sigma model} in a two dimensional curved space, with the constrain that the director $\nh$ must be tangential to the substrate.  For $K>0$,   Eq.~(\ref{F-n}) penalize the bending of the director $\nh$ in the three dimensional physical space.  {\em As a consequence, the nematic director would prefer to align with the flatter axis of principle radius, with larger radius of curvature. }  This is indeed seen in recent experiments on smectic vesicle \cite{Li-Smectic-Vesicle}.   By contrast, in the model studied by Santangelo et. al. \cite{santangelo:017801} the smectic layers prefer to align along the flatter direction.   These two models clearly describe different systems.  

The total elastic free energy for smectic on curved substrates is then given by the sum of Eq.~(\ref{F-B}) and Eq.~(\ref{F-n}), characterized by two independent parameters $B$ and $K$.

\subsection{Energetics of Toroidal Smectics}
\label{Sec:energ-torus}
Let us calculate the elastic free energy of all defects-free smectic states $(N_{\theta},N_{\phi})$ on a torus.   Strictly speaking, we have to calculate the total free energy for the generic state Eq.~(\ref{packing-torus}) with arbitrary $\Phi$ and minimize the total free energy over the elastic deformation $d\Phi(\theta,\phi)$ for given charges $(N_{\theta},N_{\phi})$.  To simplify analysis, however, we shall instead only consider states with $\Phi$ set to zero.  As we have seen in Eq.~(\ref{torus-spacing}), in the regime of thin torus $r \ll 1$,  these states have almost constant layer spacing.  Our results are not expected to be changed qualitatively by including $\Phi$ into the calculation.

The nonlinear strain Eq.~(\ref{strain-def}) can be easily calculated using Eq.~(\ref{torus-spacing}):
\begin{eqnarray}
w = 1- d_0^2 \left( n_{\theta}^2 + \frac{n_{\phi}^2}{(1+ r \cos \theta)^2} \right)
\label{strain-torus}
\end{eqnarray}
Let us first look at the infinitely thin torus limit $r \rightarrow 0$, where the strain becomes independent of $\theta$ and $\phi$.   The elastic strain energy then reduces to 
\begin{eqnarray}
F_B = \int d \omega \, \frac{B}{2} \,w^2 
\rightarrow  
\frac{1}{2} B \, A\, \left( d_0^2 n_{\theta} ^2  + 
d_0^2 n_{\phi} ^2 - 1 \right)^2,
\label{F-B-2}
\end{eqnarray} 
where $A \approx 4 \pi^2 \, R_{\theta} R_{\phi}$ is the total area of the torus.  

For arbitrary given values of $R_{\theta}$ and $R_{\phi}$, if $N_{\theta}, N_{\phi}$ were continuous variables instead of integers, we would be able to find a one-parameter family of states with vanishing strain: 
\begin{eqnarray}
w_0 &=&  d_0^2 n_{\theta} ^2  + 
d_0^2 n_{\phi} ^2 - 1
\nonumber\\
&=&  \frac{d_0^2 N_{\theta} ^2 }{(2 \pi R_{\theta})^2} 
+ \frac{d_0^2 N_{\phi} ^2 }
{(2 \pi R_{\phi})^2} - 1
\nonumber\\
& =& 0.   \label{vanish-strain-locus}
\end{eqnarray}
These ``states'' trace out an ellipse in the upper-half $(N_{\theta}, N_{\phi})$ plane~\footnote{Remember that we choose the convention $N_{\phi} \geq 0$. }, as illustrated in Fig.~\ref{strain-charges}.   In reality, of course, $(N_{\theta}, N_{\phi})$ can only take integer values.  For generic $R_{\theta}$ and $R_{\phi}$, there is no integer solution to Eq.~(\ref{vanish-strain-locus}) for $(N_{\theta}, N_{\phi})$, hence smectic state with strictly zero strain generically does not exist.   There are, however, many low energy states where the strain energy is negligibly small.  Consider the region in the $(N_{\theta}, N_{\phi})$ plane between two ellipses: 
\begin{eqnarray}
 d_0^2 n_{\theta} ^2  + 
d_0^2 n_{\phi} ^2  =  1  \pm \frac{d_0}{\sqrt{A}} , 
\end{eqnarray}
which is illustrated as the shaded region in Fig.~\ref{strain-charges}.  For all states inside this region, the absolute value of nonlinear strain is bounded by $d_0/\sqrt{A}$, hence the total strain energy Eq.~(\ref{F-B-2}) is bounded by a {\em microscopic} value: 
\begin{eqnarray}
F_B \leq \frac{1}{2} B\, d_0^2, 
\label{microscopic-strain}
\end{eqnarray}   
which is negligible comparing with all other free energy contributions, especially in the large system size regime.  The total number of these low strain energy states, i.e. the total number of grid points inside the shaded region in Fig.~\ref{strain-charges}, is roughly the total area of this low-strain-energy region in the upper-half $(N_{\theta}, N_{\phi})$ plane, which can be easily calculated:
\begin{eqnarray}
{\mathcal N}(F_B \leq  \frac{1}{2} B\, d_0^2) = 2 \pi \frac{\sqrt{A}}{d_0}.  
\end{eqnarray}
Hence {\em the total number of low strain energy states scales as the square root of the torus area.}   \footnote{Need to show that the buckling instability does not set in to the shaded region as long as the $d_0 \leq 2 \xi$, where $\xi = \sqrt{K/B}$ is the nematic penetration length.   This condition should always be satisfied.  \cite{Clark-Meyer-Smectic-Instability}. }

Hereafter we shall restrict ourself in the shaded region with low strain energy.   Let us now consider the effect of a finite but small aspect ratio $r$.   The strain energy up to the order of $r^2$ can be calculated using Eq.~(\ref{strain-torus}):
\begin{eqnarray}
F_B = \frac{1}{2} B A \left(  w_0^2 +
r^2d_0^2 n_{\phi}^2 (2 d_0^2 n_{\phi}^2 - w_0) 
\right),
\label{F-B-3}
\end{eqnarray}
where the reference strain $w_0$ is defined in Eq.~(\ref{vanish-strain-locus}).  

We have already shown that inside the shaded region in Fig.~(\ref{strain-charges}), $|w_0| \leq d_0/\sqrt{A} \ll 1$, and therefore can be ignored in Eq.~(\ref{F-B-3}). \footnote{Its inclusion, however, does not qualitatively change our results below. }  Using Eq.~(\ref{vanish-strain-locus}), then, we can parameterize $n_{\theta}$ and $n_{\phi}$ in the following way: 
\begin{eqnarray}
&& d_0\,n_{\theta}  \approx \sin \alpha 
= \frac{n_{\theta}}{\sqrt{n_{\theta}^2 + n_{\phi}^2}} ,
\nonumber\\
&& d_0\,n_{\phi} \approx \cos \alpha 
= \frac{n_{\phi}}{\sqrt{n_{\theta}^2 + n_{\phi}^2}}. 
\nonumber
\end{eqnarray}
It is not difficult to see that the geometric meaning of $\alpha$ is the angle between the nematic director $\nh$ and the local $\phi$ axis in the thin torus limit, or equivalently, the angle between the smectic layers and the local $\theta$ axis.    

The elastic strain energy Eq.~(\ref{F-B-3}) can then be simplified to 
\begin{eqnarray}
F_{\rm B} =  B A \,r^2 \left(  \cos^4 \alpha 
+ O(r) \right).   
\label{f-B-1}
\end{eqnarray}
As expected, the strain energy density scales as $r^2$ and becomes small for a thin torus.  It is in this limit that states with different global dislocation charges, as shown in the shaded region in Fig.~\ref{strain-charges}, are approximately degenerate.   
Finally, the strain energy vanishes at $\alpha = \pm \pi/2$, i.e. all smectic layers are along $\phi$ axis, and the smectic layers spacing are equal everywhere.    

The nematic distortion energy can be similarly calculated, using Eq.~(\ref{F-n}) and Eq.~(\ref{nh-psi-def}).  Again expanding to the leading order of the aspect ratio $r$, we find   
\begin{eqnarray}
F_{\nh} = \frac{2 \pi^2 K }{r} \left( \sin^2 \alpha + O(r) \right), 
\label{F-nh}
\end{eqnarray}
which clearly prefers $\alpha = 0, \pi$, for which the nematic director is parallel to the $\phi$ direction with smaller curvature.   Note, again, that $F_{\nh}$ depends on the aspect ratio $r$ but not on the substrate area $A$, hence becomes less important for larger system.   

Let us introduce a dimensionless parameter 
\begin{equation}
\chi = 2\, \sqrt{ \frac{r^3R_{\theta} R_{\phi}}{\xi^2}} 
= \frac{2\,R_{\theta}^2}{R_{\phi} \xi}
=2\, r \, \frac{R_{\theta}}{\xi}, 
\label{chi-def}
\end{equation}
where $\xi = \sqrt{K/B}$ is usually called the nematic penetration length and is of order of layer spacing $d_0$ in the strongly anchoring limit.   The total free energy, being the sum of the strain energy Eq.~(\ref{f-B-1}) and the Frank free energy Eq.~(\ref{F-nh}), is given by
\begin{eqnarray}
\frac{1}{r^2 B\, A} F_{\rm tot} &=& \cos^4 \alpha  
 + \frac{2}{ \chi^2}\,\sin^2 \alpha, 
 \label{ftot-torus}
 \\
 &=& 1+ 2 (\frac{1}{\chi^2} - 1) \alpha^2 
 + (\frac{5}{3} - \frac{2}{3 \chi^2} ) \alpha^4 + \ldots
\nonumber,
\end{eqnarray}
where we have also expanded the total free energy in terms of small angle $\alpha$.   Note that the quadratic coefficient changes sign at $\chi  = 1$.

\begin{figure} 
\begin{center}
\includegraphics[width=8cm]{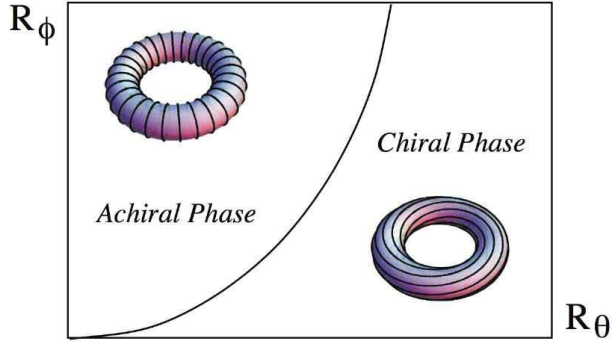}
\caption{The phase diagram of smectic on a thin torus.  The phase boundary is given by 
$R_{\theta}^2/R_{\phi} \xi = 1$.    } 
\label{phases-torus}
\end{center}
\vspace{-5mm}
\end{figure}

The ground state of smectic order on a thin torus is therefore determined by the competition between the strain energy and the Frank free energy.   Minimization of the total free energy Eq.~(\ref{ftot-torus}) over the angle $\alpha$ is straightforward.  We find two different phases.   When $\chi < 1$, the Frank free energy dominates the strain energy and the total free energy has only one minimum at $\alpha = 0$.  As illustrated in Fig.~\ref{torus-states}A, this corresponds to an achiral phase with all smectic layers forming loops in $\theta$ direction.  The nematic director is anchored by the extrinsic curvature to the local  flatter $\phi$ direction everywhere.   
In this phase, the total free energy is given by $B A \,r^2$, predominantly from the variation of layer spacing.   On the other hand, when $\chi> 1$, the total free energy has two degenerate minima at 
\begin{eqnarray}
\alpha = \pm \arccos \chi^{-1}.
\label{alpha-chi}
\end{eqnarray} 
As illustrated in Fig.~\ref{torus-states}C, this corresponds to a {\em chiral} phase with the nematic director continuously varying with the parameter $\chi$.  The spatial inversion symmetry is spontaneously broken in this phase, due to the competition between the strain energy and the Frank free energy.    Since $\alpha$ continuously increase from zero as $\chi$ increases from unity, the phase boundary is of the second order.  The phase diagram of a smectic on thin torus is summarized in Fig.~\ref{phases-torus}.  

According to Eq.~(\ref{chi-def}) if we fix the aspect ratio $r$ and let $R_{\theta}$ approach infinity, we always end up with $\chi > 1$, hence the system always falls into the chiral phase.   On the other hand, if we fix $R_{\theta}$ and let $R_{\phi}$ approach infinite, we always end up with $\chi <1$, and the system falls into the achiral phase.   This subtlety in taking the thermodynamic limit originates from the fact that the Frank free energy does not scale with the system size in the usual way.  On the other hand, for fixed system sizes $R_{\theta} \ll R_{\phi}$, if the smectic order is sufficiently weak,  the nematic penetration length may be large enough to make the parameter $\chi <1$, according to Eq.~(\ref{chi-def}).   Finally, it may be a little surprising that, according to Eq.~(\ref{alpha-chi}) and Eq.~(\ref{chi-def}), as long as the Frank constant $K$ does not vanish, the angle $\alpha$ is always {\em less than} $\pi/2$.  That is, the achiral state with $N_{\phi} = 0$ can never be the ground state.   

We must, however, be careful with the interpretation of the phase diagram Fig.~\ref{phases-torus}.   States with different $\alpha$ are separated by topological barriers.  Transitions between these states are therefore very slow.   The energetic differences between them scale sublinearly with the system size and therefore may not play significant role in physical processes.

\section{Smectic Order on Sphere}
\label{Sec:sphere}

\subsection{Spherical Nematics and Disclinations}
\label{Sec:spherical-nematics}
The first Betti number $b_1$ of a sphere is zero. Indeed, there is no non-retractable loop on a sphere: any loop on a sphere can be continuously shrunk into a single point.   Hence one would naively expect that the smectic ground state on a sphere is topologically unique.  This is, however, incorrect, due to the presence of orientational defects, i.e. disclinations.  By Eq.~(\ref{vec-psi}) a differential form $\psi$ can also be understood as a tangent vector field $\vec{\psi}$ on the sphere.   The Gauss-Bonnet-Poincar\'e theorem dictates that the total disclination charge of a tangent vector field on a sphere must equal to two.  More importantly, $\vec{\psi}$ and $-\vec{\psi}$ corresponds to the same state, hence $\vec{\psi}$ really defines a nematic order on sphere that admits half integer disclinations.   In the presence of the latter objects, we will not be able to consistently choose a direction along which the local phase field increases \footnote{Remember that $\psi$ is the differential of the local phase field of the density wave.}, and therefore $\psi$ becomes a multiple-valued differential form.  

\begin{figure} 
\begin{center}
\includegraphics[width=8cm]{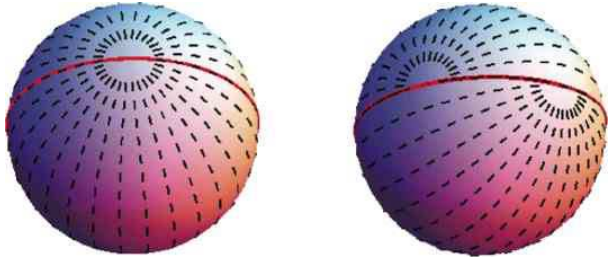}
\caption{ Generation of a one-parameter-family of degenerate ground states for spherical nematics in the limit $K_3/K_1 \rightarrow \infty$.  One first cut the sphere along a great circle that passes both poles, then rotates the two halves by an arbitrary angle.  The resulting state has four $+1/2$ disclinations on a great circle and forming a rectangle.  Note that the director field is everywhere continuous except at four disclination cores. } 
\label{sphere-nem-cut-rotate}
\end{center}
\vspace{-5mm}
\end{figure}

The ground state of spherical nematics has been subject to intensive study recently \cite{vitelli:021711,SBX-sphereical-nematics,bates:104707}, and is expected to contain four $+1/2$ disclinations in the ground state.   When the bending constant $K_3$ and splay constant $K_1$ equal to each other, four disclinations form a regular tetrahedron in the ground state.  By contrast, in the limit $K_3/K_1 \rightarrow 0 \,\,\, {\rm or} \,\,\infty$,  all bending (splay)deformations are forbidden \footnote{We note that on an arbitrary curved substrate, bending-free (splay-free) nematic texture may not exist everywhere.   The system nevertheless can exhibit domain structures.  In each domain the nematic director is bending-free (splay-free), while the director is discontinuous on boundaries between neighboring domains. }.  One can then show \cite{SBX-sphereical-nematics} that all four disclinations form a rectangle with arbitrary shape on a great circle.   The system therefore is characterized by a one-parameter family of degenerate ground states~\footnote{This degeneracy is expected to disappear as soon as the ratio $K_3/K_1$ deviates away from two fixed point, $0$ or $\infty$. }.  Interestingly, all these states can be generated from the special states with two integer disclinations at two poles, by a cut-and-rotate surgery, similar in spirit to the Volterra construction of dislocation and disclinations in solid materials.  This surgery is illustrated in Fig.~\ref{sphere-nem-cut-rotate}.   The manifold of degenerate ground states is therefore parameterized by the angle of rotation. 

It is well known \cite{LC:deGennes} that the bending constant $K_3$ (for the nematic director) is effectively infinity in the smectic phase in 3D flat space, if the layer compression modulus is large and if dislocations are energetically expensive.   
Geometrically it is easy to see that bending of director field either leads to change of layer spacing, or proliferation of dislocations.  The same result also holds in two dimensional curved space and can be elegantly derived using differential forms.  Let us write $\psi = |\psi| \nh $, where $\nh = \nh_{\alpha} dx^{\alpha}$ is the 1-form associated with the nematic director field, while $|\psi|$, the magnitude of $\psi$, is inversely proportional to smectic layer spacing, according to Eq.~(\ref{d-psi}).  Taking the exterior differential of $\psi$, as well as using the Lebiniz rule, we find 
\begin{eqnarray}
d \psi = d (|\psi| \nh) = (d|\psi|) \wedge \nh + |\psi| d \nh.
\label{dpsi-dn}
\end{eqnarray}
Clearly $d|\psi| = (\partial_{\alpha}|\psi|) dx^{\alpha}$ describes change of smectic layer spacing, while $d\nh =  (\partial_1 \nh_2 - \partial_2 \nh_1) dx^1 \wedge dx^2 $ is a 2-form describing bending deformation of the director field.  Hence if there is no dislocation, $d\psi  = 0$, and the layer spacing is constant, $d|\psi| = 0$, the bending deformation of the director field is strictly forbidden, $d\nh = 0$, which implies that the bending constant is effectively infinity.  

Conversely, starting from an arbitrary bending-free ground state for spherical nematics, it is intuitively plausible that one can glow, layer by layer, a smectic pattern with equal layer spacing and with no dislocation.  We will, however, have to fine tune the layer spacing $d$ around the energetically preferred value $d_0$ so that the smectic pattern can be fit onto the sphere with given radius $R$.  For a large sphere $R/d_0 \gg 1$, the resulting dimensionless strain Eq.~(\ref{strain-def}) should scale as $d_0/R$ \footnote{The system  may relax this strain, in a way that depends on the nematic texture. }, hence the total strain energy is bounded by  
\begin{eqnarray}
F_B \leq \frac{1}{2} B \int d \omega \, w^2 
\approx  2 \pi \, B \, d_0^2.  
\label{F-B-estimate}
\end{eqnarray}
The key feature of Eq.~(\ref{F-B-estimate}) is that {\em it does not scale with the system size}, and therefore becomes negligible compared with other energy scale as $R/d_0$ becomes large.  Furthermore, all the smectic states thus constructed naturally have equal Frank free energy.  {\em Therefore, similar to toroidal smectics, a smectic order on sphere also exhibits many approximately degenerate ground states, with the free energy difference bounded by a microscopic energy scale Eq.~(\ref{F-B-estimate}). }  Unlike the case of smectic on torus, we are not able to explicitly calculate the total free energy for each smectic state on a sphere.  \footnote{We note further that our analysis is of mean field nature, and completely ignores long wave length fluctuations at scales comparable with the substrate.  These fluctuations may further break the approximate degeneracy.  A calculation of these fluctuations is however considerably difficult and shall not be addressed in this work.} We shall calculate the total number of low energy states below.  In so far as the nematic director is concerned, therefore, a spherical smectic behaves the same as a spherical nematics in the limit of $K_3/K_1  \rightarrow \infty$, and is expected to exhibit four $+1/2$ disclinations sitting almost on a great circle \footnote{This result is probably first pointed out by Kleman and Blanc \cite{Kleman-smectic-confinement-01}. }.   This, however, does not mean that these two systems are completely equivalent.   As we shall show below, the presence of smectic layers imposes extra structure on the top of the disclination pattern, which can not be captured by the director field alone.  

\subsection{Spherical Smectic and Ambiguity in Location of Disclinations}
\label{Sec:sphere-ambiguity}

\begin{figure}
\begin{center}
\includegraphics[width=9cm]{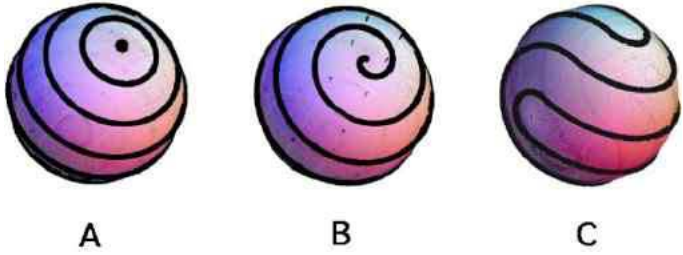}
\caption{Three types of spherical smectic states discovered in recent numerical simulations. }
\label{spherical-smectics}
\end{center}
\vspace{-5mm}
\end{figure}
Several numerical studies of spherical smectic orders in di-block copolymer films confined on sphere have been recently reported~\cite{chantawansri:lamellar-sphere,copolymer-sphere-simulation-Li,Tang-2006}.  In agreement with our discussion above, all the low energy states are found to have approximately equal layer-spacing and are free of isolated dislocation.   There may be, however, at most one layer termination (dislocation of charge one) in the disclination cores.    If there are two terminated layers near one disclination core, it saves energy to connect them.  This process is of course not elastic according to our definition in Sec.~\ref{Sec:heuristic}.  Nevertheless it happens with finite probability in the presence of thermal fluctuations, since only local rearrangement of the layer structure is needed.   We therefore infer that in all low energy smectic states on a sphere, there can be at most two open smectic layers, with possibly four layer terminations.   All other smectic layers must form close loops.  

The low energy states of smectic states are empirically classified into three categories in reference \cite{chantawansri:lamellar-sphere}: The first class is so-called hedgehog states in reference~\cite{chantawansri:lamellar-sphere} \footnote{The term hedgehog here is likely to be a misnomer.  }, shown in Fig.~\ref{spherical-smectics}A, where there are two $+1$ disclinations at north pole and south pole respectively and all layers are circles of constant latitude.   We shall however refer to them as latitudinal states.   A latitudinal state is free of layer termination everywhere.  The second class of state, as illustrated in Fig.~\ref{spherical-smectics}B, is called spiral states,  which have the same disclination pattern as the latitudinal states, but all smectic layers are spirals around the two poles.  A spiral state may have one layer termination in the center of each disclination.   The third class of state is called quasi-baseball state, as illustrated in Fig.~\ref{spherical-smectics}C, as well as Fig.~\ref{branchcuts-abcd}, where four $+1/2$ disclinations sit on a great circle and form a rectangle.  Again at most one layer termination can appear in the core of each disclination.   It is clear that each class of states contains many different states that are topologically distinct to each other.  Finally in reference \cite{chantawansri:lamellar-sphere} the total free energy of these three classes of states are also calculated; it is find that the free energy difference decreases as the system size increases, in agreement with our theoretical result discussed in the preceding subsection.  

In the remaining part of this paper, we shall explore the topological as well geometric properties of these states, as well as the connection between these states.  In particular, we shall show that both latitudinal states and spiral states can be understood as the limiting case of quasi-baseball states, where one of two dislocation charges becomes small or vanishes identically. 

\begin{figure}
\begin{center}
\includegraphics[width=6cm]{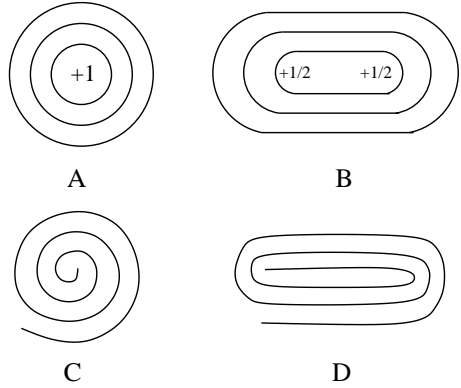}
\caption{Topologically one $+1$ disclination is not distinguishable with two $+1/2$ disclinations.  Hence A and B are topologically identical, while $C$ and $D$ are also topologically identical.   A and C are however topologically distinct to each other: There is no continuous deformation that can take state A to C.  On a sphere these continuous deformations involve significant changes of layer spacing, and therefore incur high elastic energy penalty.  
}
\label{disclination-ambiguity}
\end{center}
\vspace{-5mm}
\end{figure}

To understand the topological properties of smectic order on sphere, it is important to discuss the role played by disclinations.   Disclinations in smectic are not completely characterized by their disclination charges.  For example, states Fig.~\ref{disclination-ambiguity}A and Fig.~\ref{disclination-ambiguity}C are topologically distinct with each other, even though both have disclination charge $+1$: they can not be transformed into each other by elastic deformation.   
In fact, they can be obtained by projecting stereographically a latitudinal state and a spiral state of spherical smectic onto the complex plane.  We shall see later that state the disclination C and D should be understood as a composite defects with disclination charge $+1$ and dislocation chargs $-1$.  Composite defects with arbitrary dislocation charges $k$ can also be constructed in a straightforward way.

\begin{figure}
\begin{center}
\includegraphics[width=8cm]{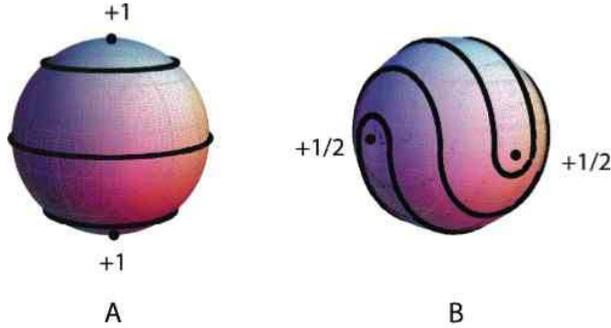}
\caption{Starting from a latitudinal state (A) with $N$ layers and with  two $+1$ disclinations at poles, we split each integer disclination into a pair of $+1/2$ disclinations, and drag them down (in perpendicular directions) to the equator.   The resulting state  (B) is a quasi-baseball state $N_1 = N_2= N$.  Only two $+1/2$ are visible in (B). }
\label{sphere-split}
\end{center}
\vspace{-5mm}
\end{figure}

It is also important to note that specification of layer positions do not completely determine the position of disclinations.  In all three states shown in Fig.~\ref{spherical-smectics}, we can only determine the location of disclinations up to the precision of layer spacing.  This problem of course also exists in flat space.   For example, for states A and B shown in Fig.~\ref{disclination-ambiguity}, we are only certain that the total disclination charge enclosed by the innermost loop is $+1$, the precise position of the disclination is however can not be determined from the layers.  
More interestingly, by continuously deforming smectic layers from state Fig.~\ref{disclination-ambiguity}A to state Fig.~\ref{disclination-ambiguity}B, we may say that we break one $+1$ disclination into a pair of $+1/2$ disclinations and separate them far apart.  There is however, no truly topological difference between these two states.   Similarly the state Fig.~\ref{disclination-ambiguity}C can also be elastically deformed into Fig.~\ref{disclination-ambiguity}D.   Return to the case of spherical smectics, such splitting and dragging of disclinations allows us to construct a quasi-baseball state from a latitudinal state by continuous deformation only.  This process is shown in  Fig.~\ref{sphere-split}.  Hence these two states Fig.~\ref{sphere-split}A and B, though seemingly very different, are really topologically identical to each other.  As we will discuss in detail below, splitting and displacement of disclinations provide the basic elastic mechanism 
that connects together many apparently different states of spherical smectics.  Nevertheless, we must note that the process of splitting and separating defects may change the layer spacing significantly, and therefore costs elastic free energy.  For example, the layer spacing in the final state shown in Fig.~\ref{sphere-split} is only half of that of the initial state.   The corresponding free energy cost is therefore proportional to the system area, and is enormous for large system, where $R/d_0 \gg 1$.

Consider a quasi-baseball state on a big sphere $R/d_0 \gg 1$, such that disclinations are well separated.   Displacement of disclinations by large distance induces significant change of layer spacing which costs high elastic free energy, and therefore is strongly suppressed even in the presence of thermal fluctuations.   Each disclination therefore fluctuates slightly around their average position.  To simplify the problem, we can treat disclinations as immobile and consider the subspace of all elastic deformations that do not change the position of any disclination, and to classify all low energy states with fixed positions of disclinations.    Two states are then called {\em weakly topologically equivalent} if they can be reached from each other by any elastic deformation that leave all disclinations fixed.   With this definition, we shall find that quasi-baseball states can also be classified by two integer global dislocation charges, much similar to the torus case.   This similarity between toroidal smectic and spherical smectic (with fixed disclinations) turns out to have interesting connection with the theory of elliptic functions, as will be shown in Sec.~\ref{Sec:sphere-quasibb}.   The stability of these states however, are not purely of topological nature.   Rather, they are stabilized by energetic as well as entropic barriers.   When large scale  motion of disclinations is allowed, many of these states actually form a continuum, mutually accessible by elastic deformtions.

By a strong contrast, if two $+1/2$ disclinations form a close bound pair, it does not cost high energy to move one relative to the other.  In particular, we can twist a disclination pair so as to exchange their positions.   This process only induces a very small change of the layer spacing (or order of $d_0/N$, where $N$ is the total number of layers), and therefore only costs small amount of elastic free energy.   Swap of disclination pairs therefore are low energy excitations in these states and should be able to be observed in simulations as well as in experiments.

In the remaining of this section, we shall study the topological as well as analytic aspects of latitudinal states and spiral states.  Analytic description of quasi-baseball states are more involved and will be discussed in Sec.~\ref{Sec:sphere}.

\subsection{Hedgehog/Latitudinal States}
\label{Sec:hedge-spiral}
Using spherical coordinate system $(\theta, \phi)$, we have a simple analytic form for latitudinal states
\begin{eqnarray}
\psi_{\rm  latitudinal} = N_{\theta} \,d \theta, 
\label{latitudinal}
\end{eqnarray}
where $\theta$ is the polar angle, and $N_{\theta}$ will be shown to be an integer below.  
The corresponding layer spacing can be calculated using Eq.~(\ref{d-psi}), with the metric tensor given by 
\begin{eqnarray}
g_{\theta\theta} = R^2, \quad g_{\phi\phi} = R^2\, \sin^2 \theta,\quad
g_{\theta\phi} = g_{\phi\theta} = 0, 
\end{eqnarray}
where $R$ is the radius of sphere.  We find that $$d = 2 \pi\,R/|N_{\theta}|. $$  Therefore Eq.~(\ref{latitudinal}) indeed describes a smectic state with equal layer spacing.  The differential form Eq.~(\ref{latitudinal}) is actually exact, with the corresponding phase field for the density wave given by:
\begin{eqnarray}
\Theta_{\rm  latitudinal} = N_{\theta} \, \theta + \Theta_0.  
\end{eqnarray}
Note, however, that $\psi_{\rm  latitudinal}$ is discontinuous at two poles,where $\theta = 0, \pi$.  The density profile nevertheless should be smooth everywhere. Using Eq.~(\ref{density-0}) we find 
\begin{eqnarray}
d \rho(\theta,\phi) &=& - \rho_1 \sin \Theta(\theta) \, d \Theta(\theta)
\nonumber\\
&=& - \rho_1 \,  N_{\theta} \, \sin \Theta(\theta) \, d \theta. 
\end{eqnarray}
which is nonsingular at both poles only if $ \sin \Theta = 0$ there. That is, 
\begin{eqnarray}
\Theta_0 = k\, \pi, \quad   N_{\theta} \pi + \Theta_0 =  k' \,\pi,  
\end{eqnarray} 
with $k$ and $k'$ are integers.   Hence  $N_{\theta}$ must be an integer.  In the setting of AB-block copolymer film, these boundary conditions imply that right at the center of disclinations, either density of specie$A$  or of specie $B$ is maximum.   

The number of smectic layers between two poles is given by 
\begin{eqnarray}
\frac{1}{2\pi}\int_N^S \psi_{\rm  latitudinal} 
= \frac{1}{2\pi} N_{\theta} \int_0^{\pi} d \theta = \frac{1}{2} N_{\theta} . 
\end{eqnarray}
Hence the geometric meaning of the integer charge $N_{\theta}$ is twice the internal distance between two poles, i.e. the number of smectic layers between them.    This internal distance is independent of the contour connecting two poles, since $\psi$ is exact. 
Now consider an infinitesimal elastic deformation generated by a vector field $\vec{X}$ that {\em preserves the locations of disclinations}, i.e. $\vec{X}$ must vanish at both poles: 
\begin{eqnarray}
\vec{X}|_{N} =  \vec{X}|_{S} = 0.  
\label{condition-NS}
\end{eqnarray}
The infinitesimal change of the differential form $\delta \psi$ induced by this elastic deformation is given by Eq.~(\ref{dela-psi-2}):
\begin{eqnarray}
\delta \psi = d \Phi = d \left( - \epsilon\, i_{\vec{X}} \psi
\right). 
\end{eqnarray}
Therefore 
\begin{eqnarray}
 \int_{N}^S \delta \psi 
&=&   \int_{N}^S d \left( - \epsilon\, i_{\vec{X}} \psi \right)
=  \left. \left( - \epsilon\, i_{\vec{X}} \psi \right)\right|^S_N,
\nonumber\\
&=& - \epsilon \, X^{\alpha} \psi_{\alpha} \left.\right|_S   
+ \epsilon \, X^{\alpha} \psi_{\alpha} \left.\right|_N
= 0,   
\end{eqnarray}
where used was Eq.~(\ref{condition-NS}).  Therefore, 
\begin{eqnarray}
\int_{N}^S \delta \psi = \int_{N}^S \delta \psi', 
\end{eqnarray}
i.e., the internal distance $N_{\theta} \pi$ between two disclinations at poles is invariant under an elastic deformation that does not displace two disclinations at poles.  $N_{\theta}$ is therefore indeed a topological quantity in the weak sense.  


\subsection{Spiral States}
\label{Sec:spiral}
In the presence of two integer disclinations at both poles, a loop that encloses the north pole once is non-retractable, as long as the loop is not allowed to cut the disclination \footnote{Consider applying arbitrary elastic deformation that leaves disclinations fixed to such a loop.  The resulting loop still encloses the north pole once. }.  The integral of $\psi$ along this loop therefore gives us an integer dislocation charge $N_{\phi}$:
\begin{eqnarray}
\oint_{\gamma_{\phi}} \psi = 2 \pi N_{\phi}.  
\label{Nphi-sphere}
\end{eqnarray}  
For the latitudinal states, this integer is clearly zero, since $\psi_{\rm  latitudinal}$ is exact.  The geometric meaning of $N_{\phi}$ is the minimal number of layers intersecting a close loop winding around the north pole.   A smectic state with one $+1$ disclinations at each pole, and with a dislocation charge $N_{\phi}$ associated with the loop $\gamma_{\phi}$, is a spiral state. 
Spiral states are therefore classified by an integer dislocation charge $N_{\phi}$.  There are two equivalent ways to understand the integer $N_{\phi}$.   If we exclude both poles from our consideration, the resulting manifold is topologically a cylinder.   It is then clear that $N_{\theta}$ is the global dislocation charge associated with the non-retractable loop winding around the cylinder once.  Equivalently, we may think of Eq.~(\ref{Nphi-sphere}) as giving the dislocation charge of the defects at the north pole.   Consequently, the defects at the north pole has both a disclination charge $+1$ and a dislocation charge $N_{\phi}$, and therefore should be understood as a composite defects.   Clearly the defects at the South pole is also a composite defects, with dislocation charge $-N_{\phi}$.   The total dislocation charge of a smectic orders on a compact substrate is always zero, according to the discussion following Eq.~(\ref{partial-T}).

We can also obtain an analytic expression for spiral states in the region that is not too close to either pole: 
\begin{eqnarray}
\psi = N_{\phi} d\phi +  a_{\theta} d \theta,  
\label{packing-sphere-spiral}
\end{eqnarray}
which clearly satisfies Eq.~(\ref{Nphi-sphere}).   Note that $a_{\theta}$ can take arbitrary real value.   
The layer spacing for this state can be calculated using Eq.~(\ref{d-psi}):
\begin{eqnarray}
d = \frac{2\pi R}{\sqrt{a_{\theta}^2 + N_{\phi}^2/\sin^2 \theta}}. 
\label{d-spiral}
\end{eqnarray}
As long as $|N_{\phi}| \ll |a_{\theta}|$ and $\sin \theta  \ll |N_{\theta}|/|a_{\theta}|$, this layer spacing is almost independent of $\theta$, hence Eq.~(\ref{packing-sphere-spiral}) is indeed a good approximation for the spiral states.  On the other hand, when $\theta \rightarrow 0, \pi$, $|\psi| \rightarrow \infty$, and the layer spacing calculated in Eq.~(\ref{d-spiral}) becomes very small.  This means that {\em Eq.~(\ref{packing-sphere-spiral})  is not a proper description of the spiral states near two poles}.  Indeed we will show below that the spiral states should really be understood as quasi-baseball states with two pairs of closely bound disclinations.  To determines smectic layer configurations in regions sufficiently close to the poles, the exact locations of two $+1/2$ disclinations around each pole are needed.  This information is however lost in the spiral state expression Eq.~(\ref{packing-sphere-spiral}), leading to singular behavior of the layer spacing in Eq.~(\ref{d-spiral}).  Finally when $a_{\theta} \sim N_{\phi}$, the description Eq.~(\ref{d-spiral}) qualitatively breaks down everywhere.   These states must be described as quasi-baseball states, which shall be studied in detail in Sec.~\ref{Sec:sphere}.

\begin{figure}
\begin{center}
\includegraphics[width=8cm]{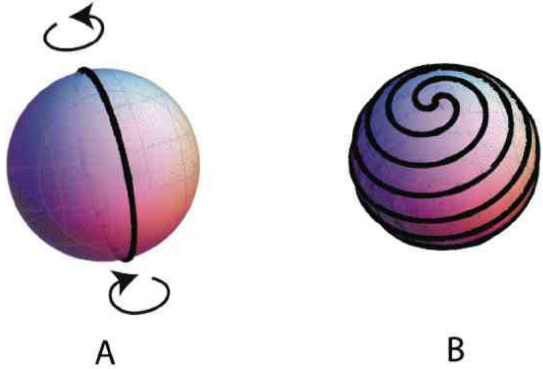}
\caption{How to elastically change the parameter $a_{\theta}$ for the spiral state Eq.~(\ref{packing-sphere-spiral}).  A) Starting from the state Eq.~(\ref{psi-spiral-proto}) with $N_{\phi} = 2$, we twist the sphere by an total angle $a_{\theta} \pi/N_{\phi}$.  B) We end up with the spiral state Eq.~(\ref{packing-sphere-spiral}).  This process also shows that the internal distance between two poles in a spiral state is not a well defined quantity, and depends on the contour over which $\psi$ is being integrated. }
\label{sphere-twist}
\end{center}
\vspace{-5mm}
\end{figure}

We have seen that in the case of latitudinal state Eq.~(\ref{latitudinal}), the parameter $N_{\theta}$ must be an integer, and is related to the internal distance between two poles.  In the case of spiral state, however, this internal distance depends on the choice of contour and is not well defined.   Therefore the corresponding parameter $a_{\theta}$ can take arbitrary real value.   In fact, we can show that spiral states with the same $N_{\phi}$ and different $a_{\theta}$ form a continuum, which are topologically identical to each other, and are mutually accessible by elastic deformations.  To see this, let us consider a prototypical state 
\begin{eqnarray}
\psi = N_{\phi} d \phi,  \label{psi-spiral-proto}
\end{eqnarray}
which describes $N_{\phi}$ smectic layers in the direction of longitudes, shown in Fig.~\ref{sphere-twist}A.  Let us elastically twist the sphere around the $z$ axis by a $\theta$ dependent angle $\Delta \phi = - \frac{a_{\theta}}{N_{\phi}} \theta$.  A point with coordinates $(\theta,\phi)$ is deformed to be the point with coordinates $(\theta',\phi')$ where 
\begin{subequations}
\label{sphere-twist-eqn}
\begin{eqnarray}
\theta \rightarrow \theta' = \theta, \quad 
\phi  \rightarrow  \phi' = \phi - \frac{a_{\theta}}{N_{\phi}} \theta.  
\end{eqnarray}
\end{subequations}
The new smectic state induced by this deformation, illustrated in Fig.~\ref{sphere-twist}B, is then given  by 
\begin{eqnarray}
\psi'(\theta',\phi') =  N_{\phi} d \phi(\theta',\phi') 
	=  N_{\phi} d\phi' + a_{\theta}d \theta' ,
\end{eqnarray}
which is precisely the spiral state Eq.~(\ref{packing-sphere-spiral}), expressed in terms of new coordinate $(\theta', \phi')$.   {\em Hence all spiral states with dislocation charge $N_{\theta}$ can be continuously generated from the prototypical state Eq.~(\ref{psi-spiral-proto}) by elastic deformations}.  
By Eqs.~(\ref{sphere-twist-eqn}) we also see that every time we twist the sphere by $\pi$  (that is, we swap the disclination pair in a counter clockwise direction), the parameter $a_{\theta}$ increases by $N_{\phi}$.   In the regime $N_{\phi} \ll a_{\theta}$, the layer spacing changes by a factor of $N_{\phi}/a_{\theta}$, with the resulting strain energy scaling as 
\begin{eqnarray}
E_B \sim B \, N_{\phi}^2 d_0^2, 
\end{eqnarray}
which only depends on $N_{\phi}$ but not on $a_{\theta}$.   Such twists of disclination pairs 
therefore costs low elastic free energy, as long as $N_{\phi}$ is of order of unity.

\section{Quasi-baseball States of Spherical Smectics}
\label{Sec:sphere-quasibb}

\begin{figure}
\begin{center}
\includegraphics[width=9cm]{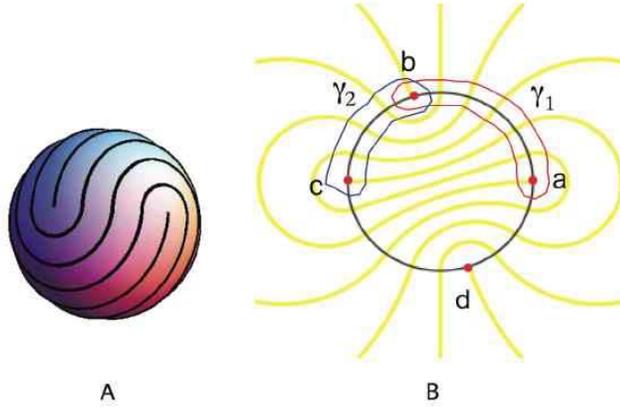}
\caption{ (Color online) (A) A quasi-baseball state, with four $+1/2$ disclinations sitting on the equator.  Each disclination core has one layer termination.  Only two disclinations are visible here.  
(B) The same state stereographically projected onto the complex plane.  The equator is mapped to the unit circle.   The red contour $\gamma_1$ encloses defects $a, b$, and intersects $12$ smectic layers in total. The blue contour $\gamma_2$ encloses defects $b, c$ and intersects $8$ layers in total.  Hence this quasi-baseball state is labeled by two integer charges $(N_1=12, N_2 = 8 )$.  }
\label{branchcuts-abcd}
\end{center}
\vspace{-5mm}
\end{figure}

\begin{figure}
\begin{center}
\includegraphics[width=6.5cm]{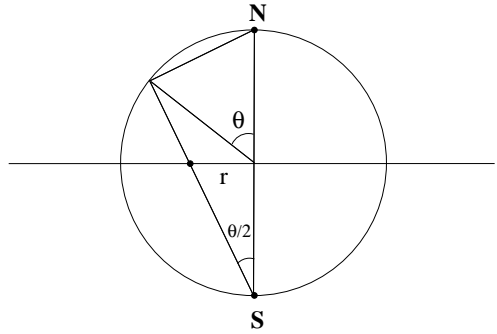}
\caption{(Side view) stereographic projection of a unit sphere onto a plane from the South pole.  The equator is projected to the unit circle. }
\label{projection}
\end{center}
\vspace{-5mm}
\end{figure}

Let us consider a quas-baseball state Fig.~\ref{branchcuts-abcd}A with four disclinations widely separated, and choose the coordinate system such that all four disclinations are on the equator.   As illustrated in Fig.~\ref{projection}, we stereographically project the sphere from the south pole to the complex plane, such that the equator is mapped into a unit circle.  The resulting smectic pattern as well as disclinations $a,b,c,d$, ordered counter clockwise, are shown in Fig.~\ref{branchcuts-abcd}B.   Let us construct a loop (red) enclosing the $ab$ segment of equator.  The minimal number of smectic layers intersecting this red loop is $N_1$.   By deforming the loop to be infinitesimally close to the segment $ab$, one clearly see that $N_1$ is precisely twice 
the total number of smectic layers intersecting the segment $ab$ of equator~\footnote{In counting this number, a layer terminating/starting right at $a$ or $b$ is counted as half.}.    Note also that because of the peculiarity of sphere, we can continuously deform the red loop, without cutting any defects, so that it tightly encloses the segment $cd$ of equator.  The integer $N_1$  can not change during such a smooth process.   Therefore $N_1$ is also twice the total number of layers cut by the segment $cd$.  Likewise, we can also introduce a (blue) loop enclosing the segment $bc$ of equator, which cut a minimal number $N_2$ of semctic layers.  A similar argument then shows that $N_2$ is also twice the total number of smectic layers intersecting the segment $bc$, as well as by the segment $da$.  We shall claim that two integers $(N_1,N_2)$ behave as global dislocation charges and completely determine the topological properties of a quasi-baseball state, under the assumption that disclinations are immobile.

Let us set up the coordinate system in the complex plane such that the four defects are given by 
\begin{eqnarray}
(a,b,c,d) = (1, e^{i \phi}, -1, e^{-i \phi}).  
\label{defects-positions}
\end{eqnarray}
From the director pattern in Fig.~\ref{sphere-nem-cut-rotate}, we easily deduce that all smectic layers intersect vertically with the equator.   Let $d$ be the equilibrium smectic layer spacing, which is uniform on the sphere, but might be slightly different with its energetically preferred value $d_0$, we find 
\begin{eqnarray}
R \, \phi &=& N_1 d,  \nonumber\\
R \, (\pi - \phi) &=& N_2 d, \nonumber
\end{eqnarray}
From which we find 
\begin{eqnarray}
\frac{N_1}{N_2} &=& \frac{\phi}{\pi - \phi}
\nonumber\\
d &=& \frac{\pi R}{N_1 + N_2}. 
\label{d-R-N}
\end{eqnarray}
The equilibrium position of defects are therefore completely determined by the two integers $N_1$ and $N_2$.  Large deviation of defects positions necessarily induces significant change of layer spacing, and therefore is energetically penalized.  

Consider a special case of quasi-baseball state with $\phi = 0$, and $N_1 = 0$.  The two $+1/2$ disclinations at $z = 1$ and $z = e^{i \phi}$ becomes degenerate and form one $+1$ disclination. 
This is clearly a latitudinal state with $N_{\theta} = N_2$, now established as a special case of quasi-baseball states.   Consider now the case $\phi \ll 1$, and $N_1 \ll N_2$.   The two $+1/2$ disclinations at $z = 1$ and $z = e^{i \phi}$ form a close bound pair, the two at $-1$ and $-e^{i \phi}$ form clearly form another pair.  If one realign the sphere such that two defects pairs are located at two poles, and calculate the integral of $\psi$ along a loop enclosing one defects pair, one finds the result $2 \pi N_1$.  This is a defining property of the spiral state with $N_{\phi} = N_1$, Eq.~(\ref{Nphi-sphere}).  Hence spiral states are also special cases of quasi-baseball states.  

Eq.~(\ref{d-R-N}) shows that $N_1+ N_2$ of every low energy states is completely determined 
by the sphere radius $R$ and the layer spacing $d$, while $N_1$ can take arbitrary integer from $0$ to $ \pi R /d$.  Hence the total number of low energy smectic states on sphere is given by 
\begin{eqnarray}
{\mathcal N}(F_B \leq  \frac{1}{2} B\, d_0^2) 
= \frac{\pi R}{d_0} + 1 
\approx \frac{\sqrt{\pi A}}{d_0},
\end{eqnarray}
i.e. scales as the square root of the total sphere area.

Our purpose is to develop an analytic description $\psi$ for low energy smectic states that have no dislocation, i.e. $d\psi = 0$ except at four disclination cores, and have equal layer spacing, i.e. $2 \pi/|\psi| = d \approx d_0$.   This turns out to be rather difficult.  Let $z =  x + i y$ is the complex coordinates obtained from the stereographic projection.  A complex differential form in the form of 
\begin{eqnarray}
\psi^c = f(z) dz
\end{eqnarray}
is called an analytic form, or holomorphic form, if $f(z)$ is an analytic function.   Using the formalism we develop in Appendix~\ref{Sec:holomorphic}, one can show that the closed differential form $\psi$ that we seek for can be expressed as the linear superposition of the real part of an analytic form $\psi^c$ and an exact form $d\alpha$:
\begin{eqnarray}
\psi(x, y) &=& Re \, \psi^c(x,y) + d \alpha(x,y), \\
\psi^c(x,y) &=&   f(z) dz , 
\end{eqnarray}
where $\alpha(x,y)$ is a smooth (but not analytic) function on the xy plane.  $\psi^H =Re \, \psi^c(x,y) $ is often called a harmonic form.   Since an exact form $d \alpha$ only describes an elastic deformation, the topological properties of the closed form $\psi$ is completely encoded in the analytic form $f(z) dz$.  In particular, as we shall show in Appendix~\ref{Sec:holomorphic}, the $+1/2$ disclinations of $\psi$ are represented as poles and branch points of the analytic function $f(z)$.  We note, however, unlike $\psi$, the layer spacing in the state described by $\psi^H$ is generically not constant.    

We have therefore broken down our task of finding the analytic descriptions of quasi-baseball states into two steps: In the first step we find a harmonic form $\psi^H$ that has the same topology as $\psi$, but does not have equal layer spacing.  In the second step we find the elastic deformation $d\alpha$ that makes the linear combination $\psi^H + d\alpha$ has equal layer spacing.  In this work, we shall only focus on the first step, and leaving the second step to an separate publication.  There is a more physical way to understand this process: one can show that $\psi^H$ describes the dislocation-free smectic ground state in the limit of vanishing layer compressional modulus $B \rightarrow 0$.   Starting from this (rather fictitious) state and adiabatically tuning up the modulus $B \rightarrow \infty$, we automatically obtain the real smectic ground state with nearly incompressible layers.

The complex analytic form $\psi^c = f(z) dz$ that describes a given distribution of disclinations is straightforward to find.  As shown in Appendix \ref{Sec:holomorphic}, the winding number of $f(z)$ at a point $z_0$ is just negative the disclination charge at the same place.  Hence a form $dz/z$ describes a $+1$ disclination at the origin, while $z dz$ describes a $-1$ disclination at the same point.  A half integer disclination at $z = a$ is naturally described by a multiple valued form $dz/\sqrt{z-a}$.   A quasi-baseball states with four $+1/2$ disclinations sitting at $a,b,c,d$ respectively is described by the following 1-form with four branch points: 
\begin{eqnarray}
\psi^c = \frac{A e^{-i\,\alpha}}{\sqrt{(z-a)(z-b)(z-c)(z-d)}} \, dz
\label{psi-abcd}
\end{eqnarray}
Our task is then to determine the complex coefficient $A\,e^{-i\alpha}$ in this holomorphic form for a given quasi-baseball state with charges $N_1$ and $N_2$.  

To avoid the ambiguity of sign associated with the multiple-valued function in Eq.~(\ref{psi-abcd}),  we introduce two branch cuts along the equator, that connects branch points $a$ and $b$, as well as $c$ and $d$ respectively, as illustrated in Fig.~\ref{branchcuts-abcd}. \footnote{There are topologically distinct ways to introduce branch cuts for a given pairing of branch points.  Exactly which one we choose is however not important; what matters is that they serve as an impenetrable fence preventing us from tracing out a close loop that winds around a single branch point.  }  The differential form Eq.~(\ref{psi-abcd}) can be made single valued, as long as it is not allowed to trespass the  branch cuts.  Furthermore, the resulting cut sphere becomes topologically indentical to a cylinder: there is only one nontractable loop, the solid loop $\gamma_1$ which encloses the branch cut $ab$ counterclockwise.  \footnote{ Note that the same loop cut also encloses the other branch cut $cd$ clockwise, since the infinity on the complex plane denotes the north pole on the sphere. }   The integral of $\psi^H$ along the close loop $\gamma_1$ defines an integer topological charge $N_1$, which we introduced at the beginning of the current section:   
\begin{eqnarray}
&& 2 \pi \, N_1 =  \oint_{\gamma_1} \psi 
= Re \, \oint_{\gamma_1} \psi^c 
,\nonumber\\
&=& Re \, \oint_{\gamma_1} 
 \frac{A e^{-i\,\alpha}  dz}{\sqrt{(z-a)(z-b)(z-c)(z-d)}}. 
\label{n1-sphere}
\end{eqnarray}
This integral is invariant under continuous deformation of the integration contour as long as it does not intersect either branch cut.  By deforming the contour sufficiently close to the branch cut $ab$, one easily see that geometric meaning of the integer charge $N_1$ is {\em twice the number of smectic layers intersecting the branch cut $ab$}.  

We can also choose another equivalent branch cuts scheme, i.e. two equator segments connecting $ac$ and $bd$ respectively, as shown in Fig.~\ref{branchcuts-abcd}.   The resulting cut-sphere, topologically a cylinder, admits one nonretractable loop $\gamma_2$, which encloses the branch cut $ac$ counterclockwise, as well as encloses $bd$ clockwise.   The integral of $\psi^c$ along $\gamma_2$ then defines the other integer topological charge $N_2$, whose geometrical meaning is twice the number of smectic layers intersecting the branch cut $ac$ (or $bd$):
\begin{eqnarray}
&& 2 \pi \, N_2 =  \oint_{\gamma_1} \psi 
= Re \, \oint_{\gamma_2} \psi^c 
\nonumber\\
&=& Re \, \oint_{\gamma_2}  
\frac{A e^{-i\,\alpha}  dz }{\sqrt{(z-a)(z-b)(z-c)(z-d)}}.  
\label{n2-sphere}
\end{eqnarray}
For given defects positions $a,b,c,d$,  two topological charges $(N_1, N_2)$ uniquely determine, through Eqs.~(\ref{n1-sphere}) and (\ref{n2-sphere}), the complex coefficient $A \, e^{-i \alpha}$ 
in the holomorphic form in Eq.~(\ref{psi-abcd}).  

\begin{figure}
\begin{center}
\includegraphics[width=5cm]{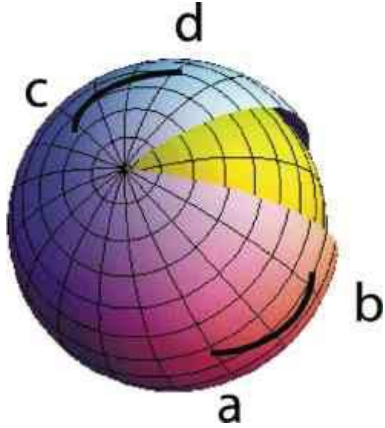}
\caption{The Riemann surface of the multiple valued function in Eq.~(\ref{psi-abcd}) consists of two concentric spheres. The outer sphere is cut open in order to show the inner one.  Both are cut along $ab$ as well as along $cd$ respectively.  They are further cross-connected through the cuts.  The resulting Riemann surface is topologically a torus, characterized by two non-retractable loops.  }
\label{Riemann-surface}
\end{center}
\vspace{-5mm}
\end{figure}

Another equiavalent, but more elegant way of representing a multi-valued complex function on the Riemann sphere is to introduce a multi-sheets Riemann surface \footnote{We refer to Stone's notes \cite{Stone-notes} for an introduction of Riemann surface and multiple valued function. }.  Each sheet is a copy Riemann sphere cut properly.  Different sheets are inter-connected appropriately in a way that is specified by the multi-valued function.  

\begin{figure}
\begin{center}
\includegraphics[width=6cm]{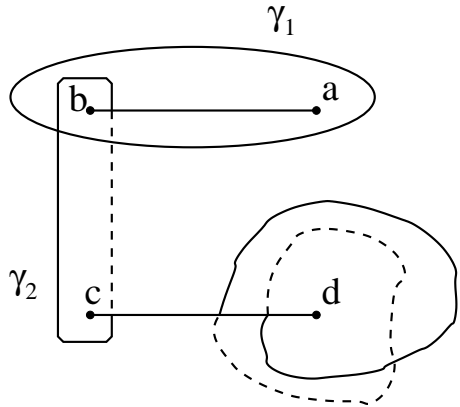}
\caption{Two independent non-retractable loops $\gamma_1$ and $\gamma_2$ on the Riemann surface shown in Fig.~\ref{Riemann-surface}. Solid lines are on the outer sheet, while dashed lines are on the inner sheet.  The loop $\gamma_1$ lies on one single sheet and encloses branch points $a$ and $b$.  $\gamma_2$ lies on both sheets and encloses branch points $b$ and $c$.   
Note that whenever a line intersect a branch cut, it leaves the sheet it resides and enters another sheet of the Riemann surface.  }
\label{Riemann-surface-loops}
\end{center}
\vspace{-5mm}
\end{figure}

The Riemann surface of the multi-valued function in Eq.~(\ref{psi-abcd}) consists of two spheres, one inside the other.  As shown in Fig.~\ref{Riemann-surface}, each of them is cut open through $ab$ and $cd$ respectively.   \footnote{Clearly we can also cut each sphere open along $ac$ and $bd$, or $ad$ and $bc$.  The topology of the final manifold obtained is however independent of this choice. }  Finally  two spheres are cross-connected along the branch cut $ab$ and along $cd$ respectively.   The topology of the Riemann surface thus obtained is however a torus~\cite{Stone-notes}, with two distinct nonretractable loops.   On of them is $\gamma_1$ as shown in Fig.~\ref{Riemann-surface-loops}, which completely lies on one sheet of the Riemann surface.    The other loop $\gamma_2$, also shown in Fig.~\ref{Riemann-surface-loops}, lies on both sheets of the Riemann surface.  The integral of $\psi^c$ along these two loops then define two dislocation charges through the same equations (\ref{n1-sphere}) and (\ref{n2-sphere}).   A curve that encloses one single branch point once is not a close curve in the Riemann surface.  The curve encloses the branch point $d$ (shown in the figure) twice is a loop that can be retracted to a single point.  One can further show that the superposition of loops $\gamma_1$ and $\gamma_2$ can be shown to be topologically identical to a loop enclosing branch points $a$ and $c$ only.

The integral
\begin{eqnarray}
w = I^{-1}(z) = \int_{z_0}^z \frac{dt} {\sqrt{(t-a)(t-b)(t-c)(t-d)}}
\label{elliptic-integral}
\end{eqnarray}
is usually called a an elliptic integral.   It defines $w$ as a multi-valued function of $z$, because the value of the integral depends on the path connecting $z_0$ and $z$.   Every time the path wraps around the non-retractable loop $\gamma_i$ once, $w$ changes by an amount $\omega_i$, given by
\begin{eqnarray}
\omega_i  = \oint_{\gamma_i} \frac{dt}{\sqrt{(t-a)(t-b)(t-c)(t-d)}}, \quad
i = 1,2. 
\label{Elliptic-periods}
\end{eqnarray}
$w = I^{-1}(z)$ is therefore only determined by $z$ up to integer multiples of $\omega_1$ and $\omega_2$.   
The inverse function $z = I(w)$ is nevertheless single-valued and is called an {\em elliptic function}, which has two periods $\omega_1$ and $\omega_2$ given by Eq.~(\ref{Elliptic-periods}):   
\begin{equation}
z(w + \omega_1) = z(w+ \omega_2) = z(w).   
\end{equation}

Remember that the physical meaning of $\psi = Re \,\psi^c$ is the differential of the local phase field of the smectic density modulation, therefore the integral of $Re \,\psi^c$ is the multiple-valued phase field $\Theta(z)$, which, according to Eqs.~(\ref{n1-sphere}-\ref{n2-sphere}), may have uncertainty $2 \pi N_1$ or $2 \pi N_2$.   Multiplying a constant $A\,e^{-i\alpha}$ to Eq.~(\ref{elliptic-integral}) and taking the real part we find 
\begin{eqnarray}
\Theta(z) = Re \int_{z_0}^z \psi^c = 
Re\,\left( A\, e^{-i\alpha} I^{-1}(z) \right).  
\end{eqnarray}   
Comparing Eq.~(\ref{Elliptic-periods}) with Eq.~(\ref{n1-sphere}) and (\ref{n2-sphere}), we find that the charges $N_i$ of the smectic state $\psi^C$ are related to the complex periods $\omega_i$ of the elliptic integral via
\begin{eqnarray}
2 \pi \, N_i = Re\, \left(  \omega_i A \,e^{-i \alpha}\right), \quad i = 1,2.  
\end{eqnarray}
This establishes the relation between the phase field $\Theta(z)$ for spherical smectic and the elliptic integral Eq.~(\ref{elliptic-integral}).

For four disclinations on equator as given by Eq.~(\ref{defects-positions}) and for given two charges $(N_1,N_2)$, let us solve Eq.~(\ref{n1-sphere}) and (\ref{n2-sphere}) for two parameters $A$ and $\alpha$.  Parameterizing $z = \exp i \varphi$ along the integrating contour, the two integrals Eqs.~(\ref{n1-sphere}-\ref{n2-sphere}) can be transformed into  
\begin{eqnarray}
\oint_{\gamma_1} \psi^c &=& i \,A\,e^{-\frac{i}{2} \phi - i \alpha} \,
\int_0^{\phi} \frac{d\varphi}{\sqrt{\sin \varphi \sin(\phi-\varphi)}}, 
\label{psi-gamma1}
\\
\oint_{\gamma_2} \psi^c &=& A\, e^{-\frac{i}{2} \phi - i \alpha} \,
\int_{\phi}^{\pi} \frac{d\varphi}{\sqrt{\sin \varphi \sin(\varphi -\phi)}}.  
\label{psi-gamma2}
\end{eqnarray}
These two integrals can be expressed in terms of complete elliptic integrals of the first kind: 
\begin{subequations}
\label{N12-result}
\begin{eqnarray}
2 \pi N_1 &=& 2 A \, K \left(\sin \frac{\phi}{2}\right) 
\sin \left(\frac{\phi}{2} + \alpha \right),  
\label{N1-result}\\
2 \pi N_2 &=& 2 A \, K \left(\cos \frac{\phi}{2}\right) 
\cos \left(\frac{\phi}{2} + \alpha \right).
\label{N2-result}
\end{eqnarray}
\end{subequations}
Derivation of these results is relegated to Appendix~\ref{App:calculation}.  
Solving Eqs.~(\ref{N12-result}) for $A$ and $\alpha$ we find
\begin{eqnarray}
A \, e^{- i \alpha} = \left( \frac{\pi N_2}{K\left( \cos \frac{\phi}{2} \right)} 
 - i \frac{\pi N_1}{K\left( \sin \frac{\phi}{2} \right) } \right)\, e^{i \phi/2},
\end{eqnarray}
where $K(\cdot)$ is the complete elliptic integral of the first kind.

\subsection{Purely Topological Classification}

\begin{figure}
\begin{center}
\includegraphics[width=7.5cm]{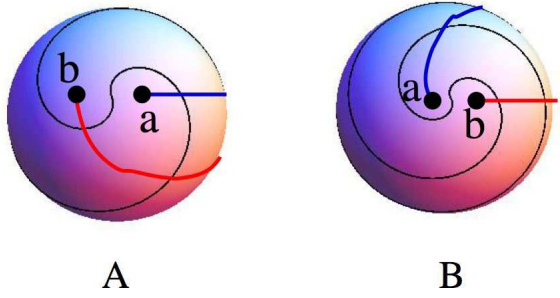}
\caption{Starting from a state $(N_1 = 2, N_2)$, we twist the sphere counter-clockwise by $\pi$ so as to exchange the defects pair $a,b$, we obtain a state $(N_1 = 2, N_2+N_1)$.  These two states are therefore topologically equivalent.  Note that this twist operation changes the layer spacing by a factor of $N_1/N_2$.  
}
\label{sphere-twist-transform}
\end{center}
\vspace{-5mm}
\end{figure}

Spiral state are especially interesting because of their closely bound defects pairs.  As illustrated in Fig.~\ref{sphere-twist-transform}, let us start with a state with charges $(N_1 = 2,N_2)$, 
fix the south pole and elastically twist the whole sphere around the z axis by an angle of $\pi$ so that the defects pair $a,b$ at the north pole exchange their positions.   Note that all smectic layers also deform with the twist ~\footnote{It is amusing to point out the similarity between this twist and a position exchange of an anyon pair. }.  The branch cut which connects $a$ and $d$ before the twist operation, shown blue in Fig.~\ref{sphere-twist-transform}A, after the twist operation, connects $b$ and $d$ intead, shown red in Fig.~\ref{sphere-twist-transform}B.  The corresponding integer charge associated with this branch cut is $N_2$ before the twist, and becomes $N_2 + N_1$  after the twist.  We therefore conclude that counter-clockwise twist of the defects pair $a,b$ leads to the following transformation of the dislocation charges 
\begin{eqnarray}
(N_1,N_2) \rightarrow (N_1,N_2 +N_1). 
\label{transform-1}
\end{eqnarray}
Obviously clockwise twist of the pair $(a,b)$ leads to the transformation of dislocation charges:
\begin{eqnarray}
(N_1,N_2) \rightarrow (N_1,N_2 - N_1). 
\label{transform-2}
\end{eqnarray}
We have then proved that, as long as defects are allowed to move, states $(N_1,N_2 \pm N_1)$ are topologically identical to the state $(N_1,N_2)$.   

More generally, we may try to twist the defects pair by an arbitrary angle $\theta$, so that the defects configuration in the twisted state does not exactly coincide with that of the initial state.  
As long as we do not focus on the smectic layer configuration near the polar regions, however, we are not able to defects the change of defects configuration.  Equivalently we are not able to unambiguously choose the branch cut that connect one of the defects near the north pole to one near the south pole.  This leads to a complete loss of the quantization of the charge $N_2$----it can take arbitrary real value.  We therefore end up with a more coarse-grained description of spiral states as discussed in Sec.~\ref{Sec:spiral}, e.g. Eq.~(\ref{packing-sphere-spiral}). 

Note, however, our twist manipulation of defects is topologically applicable for arbitrary states,
with $N_1$ or $N_2$ not necessarily small.  Starting with an arbitrary state with $N_1 \geq N_2$, and twist an appropriate defects pair so as to obtain a pair $(N'_1 = N_1 - N_2,N'_2 = N_2)$.   If $N_1' \geq N_2'$ still holds, we simply repeat the same twist operation.   If, after the twist, we find $N_2' \geq N_1'$, we twist a different pair so that we have a new pair $(N_1', N_2' - N_1')$.   This process must end after finite steps, where we end up with one charge vanishing.   Note that this reduction process is precisely the {\em Euclidean algorithm for finding the greatest common divisor $gcd(N_1,N_2)$ between two integers $(N_1,N_2)$}.  Therefore at the final state we find must have charges $(gcd(N_1,N_2),0)$, which is a latitudinal state.   Hence we have shown that 
{\em if disclinations are allowed to move, all quasibaseball states are topologically identical to some latitudinal state, with only one nonzero integer charge}.    This settles the problem of classifying all dislocation-free smectic states on sphere based on purely topological consideration.   We note however, this classification has little physical relevance, due to the significant change of layer spacing caused by twist processes.  

\begin{figure}
\begin{center}
\includegraphics[width=10cm]{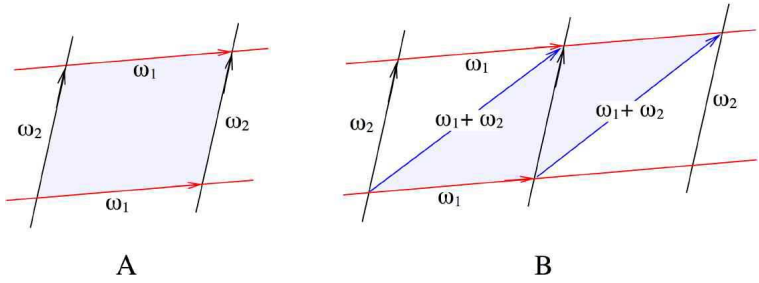}
\caption{A): A unit cell in the complex plane with bases periods $\omega_1$ and $\omega_2$ for the elliptic function $I(z)$ defined in Eq.~(\ref{elliptic-integral}), can be glued into a torus.   B): Similar to a crystal, we have freedom in choosing the two bases periods to be $(\omega_1,\omega_2)$, or to be  $(\omega_1,\omega_1+\omega_2)$.  
}
\label{two-periods}
\end{center}
\vspace{-5mm}
\end{figure}

We conclude this section with an alternative interpretation of the transformations Eqs.~(\ref{transform-1}-\ref{transform-2}).  We have already seen that Eq.~(\ref{elliptic-integral}) defines a doubly periodic function on the complex $w$ plane.   This doubly periodicity defines a lattice structure in the $w$ complex plane with a unit cell spanned by two complex numbers $\omega_1$ and $\omega_2$, as shown in Fig.~\ref{two-periods}A.  We have freedom in choosing the two periods of a doubly periodic function.  As shown in Fig.~\ref{two-periods}B, if $(\omega_1, \omega_2)$ are two periods and define a unit cell, then so are $(\omega_1, \omega_2 + \omega_1)$.  
In fact for arbitrary integers $k,l,m,n$ with $k n - lm = \pm 1$, 
\begin{eqnarray}
\omega'_1 &=& k \omega_1 + l \omega_2, \nonumber\\
\omega'_2 &=& m \omega_1 + n \omega_2, \nonumber
\end{eqnarray}
are also periods and form a unit cell.  This is because the $\omega_1$ and $\omega_2$ can also be expressed as integer combination of $\omega'_1$ and $\omega'_2$.  This redefinition of unit cell is of course well known in study of crystal structures. Twisting of a defects pair by $\pi$ therefore is equivalent to a redefinition of the unit cell in the $w$ plane for the doubly periodic function $z = I(w)$.


\section{Conclusion}
\label{Sec:conclusion}
Smectically ordered systems confined on a compact curved substrates, such as torus and sphere, exhibit rich topological as well as geometric structures.  The formalism of differential forms provides a natural framework for analyses of these structures.  

\begin{acknowledgement}
This work was supported by grant ACS PRF 44689-G7.  The author thanks M. J. Bowick for interesting discussion, as well as Monwhea Jeng and Luca Giomi for assistances on computer graphics.  
\end{acknowledgement}

\appendix

\section{Complex Differential Forms and Holomorphic Forms}
\label{Sec:holomorphic}
Our goal is to develop an analytic description in terms of differential forms for smectic states with four $+1/2$ disclinations on a sphere.  The major difficulty of this problem lies in the fact that, in the presence of $+1/2$ disclinations, it is impossible to choose a sign for the differential form $\psi$ without any ambiguity.  Parallel transport of $\psi$ (or the director field $\nh$) around one half integer disclination yields $-\psi$ (or $-\nh$).  Hence we expect that smectic orders in quasi-baseball states are described by multi-valued differential forms.   Our formalism can be applied to arbitrary curved surface, instead of sphere only.  A basic introduction of relevant formalisms can be found in chapter 7, Vol II of  M. Stone's lecture notes~\cite{Stone-notes}.  

In this section we shall exclusively use conformal coordinate system, where the metric tensor is isotropic, i.e.
\begin{eqnarray}
g_{ab} = \sqrt{g} \, \delta_{ab}, \quad
g^{ab} = \frac{1}{\sqrt{g}} \, \delta_{ab}. 
\end{eqnarray}
Note that $g$ is generically a function of coordinates.   Gauss proved that conformal coordinate system always exists for arbitrary two dimensional surface.  Let 
\begin{eqnarray}
\psi = p(x,y) dx + q(x,y) dy 
\label{psi-real}
\end{eqnarray}
be a differential form expressed in conformal coordinates $(x,y)$, the {\em Hodge dual} of the 1-form $\psi$, denoted as $\star \psi$, is defined to be  
\begin{eqnarray}
\star \psi = - q(x,y) dx + p(x,y) dy.  \label{Hodge-dual-def}
\end{eqnarray}
A more complete definition of Hodge dual using arbitrary coordinate system is given in Appendix \ref{App:Hodge-dual}.   A {\em harmonic form}  is closed form whose Hodge dual is also closed, i.e. 
\begin{eqnarray}
d \psi = d \star \psi = 0.  
\end{eqnarray} 
It is easy see that
\begin{subequations} 
\label{harmonic-form-2}
\begin{eqnarray}
d\psi &=& (\partial_x q - \partial_y p ) dx \wedge dy, \\
d \star \psi &=& (\partial_x p + \partial_y q) dx \wedge dy,
\end{eqnarray}
\end{subequations}
Therefore $\psi$ is harmonic if and only if
\begin{eqnarray}
\partial_y p = \partial_x q, \hspace{3mm}
\partial_x p = - \partial_y q,
\end{eqnarray}
which are precisely the {\em Cauchy-Riemann conditions} for a complex analytic function $f(z)= p(x,y) - i q(x,y)$.  Therefore Eq.~(\ref{psi-real}) is a harmonic form if and only if $p(x,y)$ and $-q(x,y)$ form a conjugate pair of harmonic functions.  This explains the terminology ``harmonic form''.   

If Eq.~(\ref{psi-real}) is harmonic, the complex function $f(z)= p(x,y) - i q(x,y)$ is analytic, i.e. holomorphic, where $z = x + i y$.  We can therefore introduce an {\em analytic} 1-form
\begin{eqnarray}
\psi^c =  f(z) dz = (p - i q) dz,
\label{psic-def}
\end{eqnarray} 
which depends only on $z$, but not on $\bar{z}$.   Defining the complex conjugate  $$\overline{\psi^c} =(p + i q) d\bar{z},$$ we find the real harmonic form $\psi$ and the complex holomorphic form $\psi^c$ are related to each other via  
\begin{subequations}
\label{psi-psi-c}
\begin{eqnarray}
\psi &=& \frac{1}{2} (\psi^c + \overline{\psi^c}) 
= Re \, \psi^c, \\
\psi^c &=& \psi + i \, \star \psi. 
\end{eqnarray}
\end{subequations}
The dislocation charge associated with a loop $\gamma$ is then given by
\begin{equation}
\oint_{\gamma} \psi = Re \oint_{\gamma} \psi^c.  
\end{equation}

Using the conformal metric, the magnitude of $\psi$ can be calculated easily:
\begin{eqnarray}
|\psi|^2 = g^{ab} \psi_a \psi_b = \frac{1}{\sqrt{g}} (p^2 + q^2), 
\end{eqnarray}
which is related to layer spacing by Eq.~(\ref{d-psi}).  The associated normalized 1-form (with unit magnitude) therefore equals to the nematic director form: 
\begin{eqnarray}
\nh &=& |\psi|^{-1} \psi 
= \frac{\sqrt[1/4]{g}}{\sqrt{p^2 + q^2}} (p dx + q dy)
\nonumber\\
&=& \nh_x dx + \nh_y dy. 
\label{psi-hat}
\end{eqnarray}

Let us define the magnitude of the corresponding complex form $\psi^c$ Eq.~(\ref{psic-def}) as
\begin{eqnarray}
|\psi_c|^2 = \frac{1}{\sqrt{g}} |f(z)|^2 =   \frac{1}{\sqrt{g}} (p^2 + q^2) = |\psi|^2,  
\label{abs-psi-c}
\end{eqnarray}
hence is the same as the magnitude of the real form $\psi$.  
The normalized complex 1-form is then given by 
\begin{eqnarray}
\nh^c = \frac{1}{|\psi^c|} \psi^c 
=   \frac{\sqrt[1/4]{g}}{\sqrt{p^2 + q^2}} 
	(p - i q) dz.
\label{psic-hat}
\end{eqnarray}
Comparing Eq.~(\ref{psi-hat}) with Eq.~(\ref{psic-hat}), we find that, if the smectic state is described by a harmonic form Eq.~(\ref{psi-real}), the phase of complex analytic function $f(z) = p - i q$ encodes the direction of the local nematic order relative to the conformal coordinate system $(x,y)$, that is
\begin{eqnarray}
 {\rm Arg} f(z) = - \tan^{-1} \frac{\nh_y}{\nh_x}.  \label{f-n}
\end{eqnarray}

Let us consider the nematic director Eq.~(\ref{psi-hat}) as a function of coordinate $z$.   If there is a nematic disclination at point $z_0$, the nematic director winds around the unit circle once as $z$ winds along a loop enclosing the point $z_0$ couter-clockwise.   Eq.~(\ref{f-n}) then tells us that the phase of the analytic function $f(z)$ winds the complex unit circle once clockwise.  More generally, the winding number of the analytic function $f(z)$ along a close loop is negative the disclination charge enclosed by this loop.   In other words, positive (negative) nematic disclinations are described by poles (zeros) of the meromorphic function $f(z)$.  This relation between phase of analytic function and two dimensiional orientational orders has been known for a long time and provides the basis for the application of complex analysis in nematic orders on sphere \cite{Lubensky-Prost,vitelli:021711}.  We also present a simple derivaion of this relation using differential forms in Appendix \ref{App:form-nematic}.   The basic method works, however, on arbitrary curved surface, as long as a conformal coordinate system is used.

Our purpose is however to use analytic (excluding possible singularities) function to describe {\em smectic ordering} on curved surface with no dislocation except at the disclination cores.  For this purpose we need to establish the relevance of harmonic forms, which is equivalent to holomorphic forms, as we have shown above, to smectic ordering on curved substrates.   This is achieved by the Hodge decomposition theorem of differential forms \footnote{What we have invoked here is actually only a special case of the Hodge decomposition theorem for arbitrary differential forms.}: On a compact manifold, an arbitrary closed 1-form $\psi$ can be decomposed into an exact form $d\alpha$,  and a harmonic form $\psi^H$, 
\begin{eqnarray}
\psi = d\alpha + \psi^H. 
\label{Hodge-decomp}
\end{eqnarray}
Remembering that an exact form $d \alpha$ describes an elastic deformation and does not change any topological property.   Let $\psi$ describes the smectic state that we aim to study,  the Hodge's decomposition theorem says that its topological properties can always be captured by a harmonic form $\psi^H$.  In particular, the integrals of $\psi$ and $\psi^H$ along arbitrary close loop equals to each other: 
\begin{eqnarray}
\oint_{\gamma} \psi = \oint_{\gamma} \psi^H, 
\end{eqnarray}
since the integral of an exact form $d\alpha$ along a closed loop $\gamma$ always vanishes.   Hence we conclude that in order to understand the topological structures of smectic ordering on curves surface, we only need to study harmonic forms on sphere.      

\subsection{Application to Sphere: Hedgehog States}
Let us apply the formalism of holomorphic forms to sphere.  We first stereographically project a unit sphere into a plane from the south pole.  As illustrated in Fig.~\ref{projection}, an arbitrary point $(\theta, \phi)$ on the sphere is projected to the point $(r = \tan \theta /2, \phi)$ using polar coordinate system.  The north pole is projected to the origin, the south pole to the infinity, while the equator $\theta = \pi/2$ to the unit circle $r = 1$.  Introducing the Cartesian coordinate $(x,y)$ in the plane:
\begin{eqnarray}
x = r \,\cos \phi, \quad y = r \, \sin \phi,
\end{eqnarray}
we easily verify that they are indeed conformal coordinates:
\begin{equation}
g_{xx} = g_{yy} = \frac{1}{(1+r^2)^2},
\quad
g_{xy} = 0.  
\label{metric-sphere}
\end{equation}
We can then subsequently form a complex coordinate $z = x + i y$.

Let us first consider the holomorphic 1-form 
\begin{eqnarray}
\psi^c = f(z) dz = \frac{1}{z} dz.  
\label{psic-sphere-1}
\end{eqnarray} 
The winding number of $\frac{1}{z} $ around $z_0 = 0$ is clearly $-1$, hence Eq.~(\ref{psic-sphere-1}) describes a $ +1$ nematic disclination at $z = 0$, i.e. the north pole, according to the stereographic projection.  In order to look at the configuration around the south pole, we should project from the south pole.  The resulting new complex coordinate (also conformal) $w$ is related to $z$ via $w= 1/z$.  In terms of $w$ the 1-form $\psi^c$ is given by
\begin{eqnarray}
\psi^c = \frac{1}{z}\, dz = - \frac{1}{w} \, dw.  
\end{eqnarray} 
Hence the winding number around $w = 0$ ($z = \infty$) is also $-1$, which corresponds to a $+1$ nematic disclination at south pole.  This is of course rather expected, as the total disclination charge has to be $+2$ by the Gauss-Bonet-Poincar\,e theorem.   

Let us look at the corresponding real 1-form, using polar expression $z = r\, e^{i \,\phi}$ as well as the relation Eq.~(\ref{psi-psi-c}) we find 
\begin{eqnarray}
\psi = Re \, \psi^c = \frac{1}{r} \,d r = d\left( \log \, r \right)
= d \left( \log \tan \frac{\theta}{2} \right),
\end{eqnarray}
which is actually an exact form.   The layer spacing can also be calculated, using Eq.~(\ref{d-psi}) and Eq.~(\ref{abs-psi-c}), as well as Eq.~(\ref{metric-sphere}), 
\begin{equation}
d = \frac{2 \pi}{|\psi|} = \frac{2 \pi\, r}{1+r^2}, 
\end{equation}
which approaches zero as $r \rightarrow 0$ (north pole) and $r \rightarrow \infty$ (South pole).  Hence smectic layers become infinitely dense around both poles.  This peculiarity is a consequence of $\psi$ being harmonic.  As it stands the harmonic form $\psi$ does not describe the ground state with minimal energy.  The ground state which clearly prefers equal layer spacing, should be solved by using Eq.~(\ref{Hodge-decomp}) and minimizing the total elastic free energy over the elastic deformation $d \alpha$.   Such an analysis is left to a separate publication.   

Let us integrate $\psi$ and find the phase field $\Theta$: 
\begin{equation} 
\Theta(\theta) - \Theta_0
= \log \frac{\tan \frac{\theta}{2}}{\tan \frac{\theta_0}{2}}
\end{equation}
where $\theta_0$ and $\Theta_0$ are two arbitrary constants.  Hence all smectic layers form circles of constant latitude, while nematic director is parallel to lines of longitude. Hence Eq.~(\ref{psic-sphere-1}) describes a latitudinal state.   This result can also be seen by calculating the dislocation charge associated with the loop $\gamma_{\phi}$: 
\begin{equation}
\oint_{\gamma_{\phi}} \psi 
= Re \oint_{\gamma_{\phi}} \psi^c = 0, 
\end{equation}
which corresponds to a latitudinal state.  

Finally we note that description of smectic layers in terms of 1-form Eq.~(\ref{psic-sphere-1}) only works outside the disclination core regions.

\subsection{Spiral States and Composite Defects}
Let us now consider a slightly more complicated 1-form:
\begin{eqnarray}
\psi^c = f(z) dz = A \,e^{-i \alpha}\,\frac{1}{z} dz.  
\label{psi-spiral}
\end{eqnarray}
It still describes two $+1$ disclinations sitting on two poles.  On the other hand, the real 1-form is given by 
\begin{eqnarray}
\psi = Re(\psi^c) = A \,\cos\alpha \, \frac{1}{r} dr + A\,\sin\alpha \, d \phi,
\end{eqnarray}
where $\phi$ is the polar angle of the spherical coordinate system.  Integrating $\psi$ along a loop $\gamma_{\phi}$ enclosing north pole once, one must obtain an integer multiple of $2 \pi$, if layers are well defined.  That is, 
\begin{eqnarray}
\oint_{\gamma_{\phi}} \psi = 2 \pi\, A\,\sin\alpha  = 2 k \pi,
\end{eqnarray}
which imposes constraint on the parameters $A$ and $\alpha$.   Eq.~(\ref{psi-spiral}) therefore describes a spiral state with global dislocation charge (integer) given by $A\,\sin\alpha$.  The disclinations at both poles therefore are composite defects with both disclination charge and dislocation charge.  It is rather interesting to see that these composite defects Eq.~(\ref{psi-spiral}) can be generated from pure disclination Eq.~(\ref{psic-sphere-1}) by simple multiplication of a phase factor.  

It is also easy to find the locus of constant phase $\phi_0$ of density wave by integration:
\begin{eqnarray}
\log r + \phi \,\tan \alpha = \frac{\phi_0}{A},
\end{eqnarray}
which is a {\em logarithmic spiral}.  Again we note that description in terms of differential form only works outside the disclination cores, i.e. $ \tan^{-1}a_0 < r < \tan^{-1} 1/a_0$.  

More generally, let $\psi^c = f(z) dz$ be a holomorphic (except a few poles or branch cuts) form describing some particular smectic state with no dislocation.  The integral of $\psi^c$ along a close contour $\gamma$ is generically a complex number: 
\begin{eqnarray}
\oint_{\gamma} \psi^c = \oint_C f(z) dz = B_R + i \, B_I.  
\label{psic-integral}
\label{oint-psic}
\end{eqnarray}
Using Eq.~(\ref{psi-psi-c}) we find that 
\begin{eqnarray}
\oint_{\gamma} \psi = B_R, \quad
\oint_{\gamma} \star \psi = B_I.  
\end{eqnarray}
That is, $B_R$ and $B_I$ are the dislocation charges associated with the loop $\gamma$ 
for smectic states $\psi$ and $\star \psi$ respectively.   We note that Eq.~(\ref{oint-psic}) is the sum of residue of $\psi = f(z) dz$ enlcosed by the loop.

Let us apply a global rotation of the form $\psi^c$ by multiplying a complex phase factor:
\begin{eqnarray}
\psi^c \rightarrow \psi^{c'} = e^{- i \alpha} \psi^c.  
\end{eqnarray}
The new real differential from $\psi'$ is then given by
\begin{eqnarray}
\psi' = Re \, \psi^{c'} =  \cos \alpha \, \psi + \sin \alpha \, \star \psi,
\end{eqnarray}
where $\star \psi$ is the Hodge dual of $\psi$, defined by Eq.~(\ref{Hodge-dual-def}).  
The dislocation charge associated with $\psi'$ is then given by
\begin{eqnarray}
\oint_C \psi' = Re \, \oint_C \psi^{c'} 
= B_R \cos \alpha + B_I \, \sin \alpha, 
\label{Br-BI}
\end{eqnarray}
which is generically different from $B_R$.  Note that $|\psi^c| = |{\psi^c}'|$, which means that layer spacing are the same in two states $\psi^c$ and ${\psi^c}'$.  
 Multiplication of a complex phase factor to a holomorphic form therefore provides a basic mechanism for changing dislocation charge without changing layer spacing.   Note howeve Eq.~(\ref{Br-BI}) must be integer multiple of $2 \pi$.  This means that for given value of $B_R$ and $B_I$, only discrete choices of the angle $\alpha$ are compatible with smectic ordering.

\section{Calculation of Integrals Eqs.~(\ref{psi-gamma1}-\ref{psi-gamma2})}
\label{App:calculation}

To calculate the integral in Eq.~(\ref{psi-gamma1}), we use the identity 
\begin{eqnarray}
\sin(\varphi) \sin (\phi - \varphi) = 
\frac{1}{2} \cos(2 \varphi - \phi)
-\frac{1}{2} \cos \phi
\end{eqnarray}
and introduce a new integral variable $\beta = 2 \varphi - \phi$.  The original integral is then transformed to 
\begin{eqnarray}
\oint_{\gamma_1} \psi^c 
&=& \frac{iA}{\sqrt{2}} e^{-\frac{i}{2} \phi - i\alpha}
\int_{-\phi}^{\phi} \frac{d\beta}{\sqrt{\cos \beta - \cos\phi}}
\nonumber\\
&=&  \frac{iA}{2} e^{-\frac{i}{2} \phi - i\alpha}
\int_{-\phi}^{\phi} \frac{d\beta}
{\sqrt{\sin ^2 \frac{\phi}{2} - \sin^2 \frac{\beta}{2}}}.
\label{int-psi-1}
\end{eqnarray}
Introducing a new variable $t$ through
\begin{eqnarray}
t = \frac{\sin \beta / 2}{\sin \phi/2}, 
\end{eqnarray}
we transform the integral Eq.~(\ref{int-psi-1}) into a complete elliptic integral of the first kind:
\begin{eqnarray}
\oint_{\gamma_1} \psi^c 
&=& 2 i A e^{-\frac{i}{2} \phi - i\alpha} 
\int_0^1 \frac{dt}{\sqrt{(1-t^2)(1-t^2 \sin^2\frac{\phi}{2})}}
\nonumber\\
&=& 
2 i \,A\,e^{-\frac{i}{2} \phi - i\alpha} K \left(\sin \frac{\phi}{2}\right).  
\label{int-psi-1-1}
\end{eqnarray}

To calculate integral Eq.~(\ref{psi-gamma2}), we make variable transformations:
\begin{eqnarray}
\bar{\varphi} = \pi - \varphi, \,\,\,\,
\bar{\phi} = \pi - \phi.
\end{eqnarray}
Eq.~(\ref{psi-gamma2}) then reduces to 
\begin{eqnarray}
\oint_{\gamma_2} \psi^c   = 
A \, e^{-\frac{i}{2} \phi- i \alpha} \,
\int_{0}^{\bar{\phi}} \frac{d\bar{\varphi}}
{\sqrt{\sin \bar{\varphi} \sin (\bar{\phi} - \bar{\varphi})}},
\end{eqnarray}
which as the same functional form as Eq.~(\ref{psi-gamma1}).
Now using the result Eq.~(\ref{int-psi-1-1}) we find 
\begin{eqnarray}
\oint_{\gamma_2} \psi^c  &=& 
2 \,A \,e^{ - \frac{i}{2} \phi - i\alpha} K \left(\sin \frac{\bar{\phi}}{2} \right)
\nonumber\\
 &=&  2 \,A \,e^{ - \frac{i}{2} \phi - i\alpha} K \left(\cos \frac{\phi}{2} \right). 
\label{int-psi-2-1}
\end{eqnarray}

Combining Eqs.~(\ref{int-psi-1-1}) and (\ref{int-psi-2-1}) with Eqs.~(\ref{n1-sphere}-\ref{n2-sphere}) we finally obtained two equations that uniquely determine $A$ and $\alpha$:
\begin{eqnarray}
2 \pi N_1 &=& 2 A \, K \left(\sin \frac{\phi}{2}\right) 
\sin \left(\frac{\phi}{2} + \alpha \right)
,\\
2 \pi N_2 &=& 2 A \, K \left(\cos \frac{\phi}{2}\right) 
\cos \left(\frac{\phi}{2} + \alpha \right)
.
\end{eqnarray}

\section{Hodge Dual and Codifferential}
\label{App:Hodge-dual}
In this appendix, we list the relevant results about Hodge dual of differential forms as well as codifferential operator.  All these results can be found in standard textbooks of differential geometry, see, for example \cite{book:Nakahara,book:Frankel,Stone-notes}.   

In a $d$ dimensional manifold, the Hodge dual of a p-form is a $(d-p)$-form defined as  
\begin{eqnarray}
\star dx^{i_1}\wedge \cdots \wedge dx^{i_p} &=& \frac{1}{(d-p)!}
\sqrt{g} \,g^{i_1j_1}\cdots g^{i_p j_p}
\\
&& \epsilon_{j_1\cdots j_p j_{p+1}\cdots j_d} 
dx^{j_{p+1}} \cdots dx^{j_d}. 
\nonumber
\end{eqnarray}
For a two dimensional Riemannian manifold with orthogonal coordinate system $(x,y)$, we have 
\begin{subequations}
\label{Hodgedual}
\begin{eqnarray}
\star 1 &=& \sqrt{g} \, dx \wedge dy,\\
\star dx &=& \sqrt{g} \, g^{xx} dy,\\
\star dy &=& - \sqrt{g} \, g^{yy} dx,\\
\star dx \wedge dy &=& \frac{1}{\sqrt{g}}. 
\end{eqnarray}
\end{subequations}
It is clear that when acting on a $p$-form twice, we have 
\begin{eqnarray}
\star \star \omega^p = (-1)^{p(d-p)} \omega^p. 
\end{eqnarray} 
For a conformal coordinate system, $g^{xx} = g^{yy} = 1/\sqrt{g}$, and Eqs.~(\ref{Hodgedual}) reduces to 
\begin{subequations}
\label{Hodgedual-conformal}
\begin{eqnarray}
\star 1 &=& \sqrt{g} \, dx \wedge dy,\\
\star dx &=& dy,\\
\star dy &=& -  dx,\\
\star dx \wedge dy &=& \frac{1}{\sqrt{g}}. 
\end{eqnarray}
\end{subequations}

If $\alpha$ and $\beta$ are two $p$-form fields, $a \wedge \star b$ is a $d$-form field, which can be integrated over the whole manifold.  This allows us to define an inner-product of two $p$-form fields
\begin{eqnarray}
\langle \alpha, \beta \rangle &=& \langle \beta,\alpha \rangle = \int_{\Omega} \alpha \wedge \star\beta
\nonumber\\
&=& \frac{1}{p!} \int d\omega \, \alpha_{i_1\cdots i_p} \beta^{i_1\cdots i_p},
\label{form-inner-prod}
\end{eqnarray}
where in the last equality, the indices of $\beta$ have been raised using the metric tensor.  

The codifferential operator $\delta$ is defined as 
\begin{eqnarray}
\delta &=& (-1)^{dp + d + 1} \star d \star 
\\
&\rightarrow& - \star d \star 
, \quad {\rm for} \,\,  d= 2 . 
\nonumber
\end{eqnarray}
Note that $\delta$ maps a $p$ form into a $(p-1)$-form, i.e. it works in the opposite direction as $d$. 
$\delta$ acting on $0$-form is defined to be zero.  It can be shown that $\delta$ is the adjoint of $d$, i.e. 
\begin{eqnarray}
\langle \alpha,d\beta \rangle = \langle \delta \alpha, \beta \rangle.  
\label{delta-d}
\end{eqnarray}
One can also show that 
\begin{eqnarray}
\delta^2 = d^2 = 0. 
\end{eqnarray}

The self-adjoint Laplace-Beltrami operator acting on differential form is defined as
\begin{eqnarray}
\Delta \alpha = (\delta d + d \delta) \alpha.
\end{eqnarray}
One can show that 
\begin{eqnarray}
\langle \alpha, \Delta \beta \rangle = \langle \Delta \alpha, \beta \rangle 
= \langle d\alpha, d\beta \rangle + \langle \delta \alpha, \delta \beta \rangle. 
\nonumber\\ 
\end{eqnarray}
Setting $\alpha = \beta$, one find 
\begin{eqnarray}
\langle \alpha, \Delta \alpha \rangle = 
\langle d \alpha,  d \alpha \rangle 
+ \langle \delta \alpha,  \delta \alpha \rangle  
\geq 0. 
\end{eqnarray}
Hence the Laplace-Beltrami operator must have non-negative eigenvalues.

\section{Differential Form and Nematics on Curved Substrates}
\label{App:form-nematic}
In this appendix we present a simple by elegant derivation of the connection between equilibrium nematic texture on curved substrates and holomorphic functions using differential forms.   Let $D_{\alpha}$ denote the covariant derivative associated with the symmetric Levi-Civita connection of the manifold, we easily see  
\begin{eqnarray}
D_{\alpha}\nh_{\beta} - D_{\beta}\nh_{\alpha}
 = \partial_{\alpha}\nh_{\beta} - \partial_{\beta}\nh_{\alpha}
\end{eqnarray}
Consider the differential form $\nh = \nh_{\alpha} dx^{\alpha}$ describing the nematic director field,
and for simplicity using a conformal coordinate system. Using Eq.~(\ref{form-inner-prod}) we find
\begin{eqnarray}
d \nh &=& (\partial_1 \nh_2 - \partial_2\nh_1) dx^1\wedge dx^2,\\
\langle d\nh,d\nh \rangle &=& 
\int d\omega \, g^{-1} (\partial_1 \nh_2 - \partial_2\nh_1)^2.  
\end{eqnarray}
We recognize that this is precisely the in-plane bending component of the Frank free energy for nematic director in curved space.  On the other hand, we also have
\begin{eqnarray}
\delta \nh &=& \star d \star (\nh_1 dx^1 + \nh_2 dx^2) 
\nonumber\\
&=& \star d (\nh_1 dx^2 - \nh_2 dx^1)
\nonumber\\
&=& \star \left(
\partial_1 \nh_1 + \partial_2 \nh_2
 \right) dx^1 \wedge dx^2 
 \nonumber\\
 &=& \frac{1}{\sqrt{g}}  \left(
\partial_1 \nh_1 + \partial_2 \nh_2
 \right) 
\end{eqnarray}
One can then show that 
\begin{eqnarray}
\langle \delta \nh, \delta \nh \rangle = 
\int d\omega (D_{\alpha} \nh^{\alpha})^2,  
\end{eqnarray}
is the splay component of the Frank free energy. 

The intrinsic part \footnote{as supposed to the extrinsic part of free energy that involving the extrinsic curvature} of the Frank free energy for nematic order on curved substrate then can be expressed in terms of differential form as 
\begin{eqnarray}
F_{\nh} =\frac{K_1}{2} \langle \delta \nh, \delta \nh \rangle  + 
\frac{K_3}{2}  \langle d\nh,d\nh \rangle. 
\end{eqnarray}
When $K_1 = K_3$, this can be further simplified using the Laplace-Beltrami operator: 
\begin{eqnarray}
F_{\nh} = \frac{K}{2} \langle \nh, \Delta \nh \rangle.  
\end{eqnarray}

Minimizing the Frank free energy over  $\nh$ subject to the constraint $|\nh| = 1$, one find the Euler-Lagrange equation for $\nh$:
\begin{eqnarray}
\Delta \nh  + \lambda(x) \, \nh = (\delta d + d \delta) \nh + \lambda(x) \, \nh = 0,
\label{nh-eqn}
\end{eqnarray}
where $\lambda(x)$ is a Lagrange multiplier field which generically depends on $x$. 
Let the function (0-form) $\phi(x)$ satisfy the same equation as $\nh$, i.e.
\begin{eqnarray}
\Delta \phi + \lambda \,\phi = \delta d \, \phi (x) + \lambda \phi(x) = 0.  
\end{eqnarray}
we can show that the 1-form $\psi = \phi \, \nh$ is a harmonic form, i.e. 
\begin{eqnarray}
\Delta \psi = (\delta d + d \delta ) \psi = 0.  
\end{eqnarray}
Hence we have 
\begin{eqnarray}
\nh(x) = \frac{1}{\phi(x)} \psi(x).
\end{eqnarray}

Conversely, if $\psi(x)$ is a harmonic form, and let us define $\nh(x) = \psi(x)/|\psi(x)|$ (hence $\nh(x)$ is properly normalized),
and $\lambda(x) = - \delta d |\psi(x)|/|\psi(x)|$, then one can show that $\nh(x)$ automatically satisfies Eq.~(\ref{nh-eqn}), i.e. it indeed describes an equilibrium nematic texture.  

This establishes the connection between the equilibrium nematic texture and harmonic form $\psi(x)$.  Using the relation between harmonic form and holomorphic form, we further obtain the (well known) relation between equilibrium nematic texture and holomorphic functions.


\end{document}